\documentclass[12pt]{article}
\usepackage{amsmath, amsthm, amssymb, bbm}
\usepackage{graphicx, subcaption, caption, float, adjustbox, psfrag, epsf, color}
\usepackage{setspace, enumerate, hyperref, xcolor, arydshln}
\usepackage[ruled,vlined]{algorithm2e}
\usepackage[
  backend=bibtex,
  style=authoryear,
  citestyle=authoryear,
  maxbibnames=99
]{biblatex}
\newcommand{\blind}{0}

\theoremstyle{definition}
\newtheorem{definition}{Definition}

\begin{document}

\def\spacingset#1{\renewcommand{\baselinestretch}%
{#1}\small\normalsize} \spacingset{1}

\if0\blind
{
  \title{\bf Population Quasi-Monte Carlo}
  \author{\\
    Chaofan Huang \hspace{1cm} V. Roshan Joseph \\
    H. Milton Stewart School of Industrial and Systems Engineering, \\
    Gerogia Institue of Technology, Atlanta, GA, 30332 \\
    \\
    and \\
    \\
    Simon Mak \\
    Department of Statistical Science, \\
    Duke University, Durham, NC, 27708}
  \maketitle
} \fi

\bigskip
\begin{abstract}
Monte Carlo methods are widely used for approximating complicated, multidimensional integrals for Bayesian inference. Population Monte Carlo (PMC) is an important class of Monte Carlo methods, which utilizes a population of proposals to generate weighted samples that approximate the target distribution. The generic PMC framework iterates over three steps: samples are simulated from a set of proposals, weights are assigned to such samples to correct for mismatch between the proposal and target distributions, and the proposals are then adapted via resampling from the weighted samples. When the target distribution is expensive to evaluate, the PMC has its computational limitation since the convergence rate is $\mathcal{O}(N^{-1/2})$. To address this, we propose in this paper a new \textit{Population Quasi-Monte Carlo} (PQMC) framework, which integrates Quasi-Monte Carlo ideas within the sampling and adaptation steps of PMC. A key novelty in PQMC is the idea of importance support points resampling, a deterministic method for finding an ``optimal'' subsample from the weighted proposal samples. Moreover, within the PQMC framework, we develop an efficient covariance adaptation strategy for multivariate normal proposals. Lastly, a new set of correction weights is introduced for the weighted PMC estimator to improve the efficiency from the standard PMC estimator. We demonstrate the improved empirical convergence of PQMC over PMC in extensive numerical simulations and a friction drilling application.
\end{abstract}

\noindent%

\newpage

\spacingset{1.45}

\section{Introduction}
\label{sec:introduction}

A fundamental challenge in Bayesian inference is the evaluation of integrals involving some multi-dimensional posterior distribution $\pi$. Generally, closed-form analytical solutions are not feasible, and Monte Carlo (MC) methods are often used for approximation. Of such methods, Markov Chain Monte Carlo (MCMC; \cite{robert2013mc}) is widely used due to ease of implementation. In recent decades, there has been renewed interest in exploring an alternative class of methods called iterated Importance Sampling (IS; \cite{robert2013mc}), which allows for parallel implementation, flexibility of adaptation, and easy assessment of approximation error over MCMC. However, the success of iterated IS depends on finding a good set of proposal distributions that mimics the target distribution $\pi$, which can be difficult when $\pi$ is high-dimensional and/or time-consuming if $\pi$ is computationally expensive to evaluate. To address this, we propose a novel Population Quasi-Monte Carlo (PQMC) framework, which integrates Quasi-Monte Carlo sampling within the Population Monte Carlo (PMC; \cite{cappe2004pmc}) -- a popular iterated IS method -- for improved sampling performance. \par


The key idea in PMC is to adapt a population of proposals iteratively, to generate weighted samples which are approximately drawn from $\pi$. This adaptation idea can be traced back to \textcite{oh1993ais}, \textcite{west1993ais}, and \textcite{givens1996lais}. At each iteration, the PMC algorithm first simulates $J$ \textit{samples} from each of the $K$ proposal distributions $\{q_{k}\}_{k=1}^{K}$, i.e. $x_{k,j}\sim q_{k}$. Next, it \textit{weighs} the obtained $KJ$ samples $\{x_{k,j}\}_{k=1}^{K}{}_{j=1}^{J}$ to correct for mismatch between the proposal and target distributions. Last, it updates the $K$ proposals via \textit{resampling}, so that samples with larger weights are duplicated and samples with insignificant weights are eliminated, thereby allocating more resources for exploring higher probability regions. These \textit{sampling, weighting, adaptation} steps are then repeated for $T$ iterations, yielding a total of $N = TKJ$ weighted samples for approximating $\pi$. We note that there are other adaptation variants of PMC which do not rely on resampling, such as D-kernel PMC \parencite{douc2007dmis1,douc2007dmis2}, Mixture PMC \parencite{cappe2008mmis}, Adaptive Population Importance Sampler \parencite{martino2015apis}, and Random Walk Importance Sampler \parencite{martino2017lais}. In this paper, we only focus on the PMC with resampling adaptation scheme, which enjoys the same convergence rate of $\mathcal{O}(N^{-1/2})$ as IS. \par

In the literature, there are two main \textit{weighting} strategies which both yield unbiased integral estimates: the standard importance weights $w(x_{k,j}) = \pi(x_{k,j})/q_{k}(x_{k,j})$, and the deterministic mixture weights $w(x_{k,j}) = \pi(x_{k,j})/[K^{-1}\sum_{i=1}^{K}q_{i}(x_{k,j})]$. \textcite{elvira2019mis} proved theoretically that the latter mixture weighting scheme has smaller variance for integral estimation. However, this mixture weighting strategy requires $\mathcal{O}(K^2J)$ evaluations of the proposal distributions. When $K$ is large, this is a major computational burden compared to the standard weighting strategy where only $\mathcal{O}(KJ)$ evaluations are required. One way to reduce $K$ while keeping the total number of samples $N$ fixed is to increase $J$. A large $J$, e.g. $J = 10$, was also proposed as a remedy to the sample impoverishment issue  in resampling, i.e., it is possible for the samples to collapse to only a few particles with very large weights \parencite{carpenter1999sr}. \textcite{elvira2017dmmis} also shows empirical improvement for PMC when $J > 1$ under the deterministic mixture weighting scheme. 

However, with a smaller number of proposals $K$, we could lose too much information when down-sampling the $KJ$ simulated samples to $K$ particles via resampling. One solution is to incorporate Quasi-Monte Carlo (QMC; \cite{niederreiter1992random}) into the \textit{resampling} step. QMC uses a set of low discrepancy deterministic points which are well spread out over the sample space to achieve better convergence rate for integration. Thus, by obtaining $K$ ``space-filling'' (i.e., well spaced-out) points that retain the most information from the $KJ$ simulated samples via QMC resampling, we can reduce the additional Monte Carlo error introduced in the \textit{resampling} step. Moreover, the use of ``space-filling" resamples as the new proposals enables more efficient exploration of the parameter space. For this, we propose a new deterministic resampling method called importance support points (ISP) resampling, which makes use of the support points in \textcite{mak2018sp} to find the set of resamples which ``best'' represents the weighted proposal samples. \par

Moreover, it is known that the QMC convergence rate can achieve $\mathcal{O}(N^{-1}(\log N)^{p-1})$ for integration in uniform hypercube, where $p$ is the dimension of the parameters \parencite{owen2013mc}. Hence, it is also beneficial to leverage QMC in the \textit{sampling} step by using low-discrepancy samples, leading to more representative points from each proposal distribution. With QMC sampling, we can also reduce $J$, the number of samples simulated from each proposal, and thus reducing $N$, the total number of posterior evaluations, while still achieving the desired precision. With the above two QMC modifications to the sampling and resampling steps of the PMC algorithm, we propose a novel Population Quasi-Monte Carlo (PQMC) framework that provides significant improvement over the PMC algorithm. The faster empirical convergence of PQMC makes it a useful tool for efficiently sampling from posterior distributions which are computationally expensive; such posteriors often arise in complex engineering applications \parencite{joseph2019mined}. In recent years, QMC has been adapted for speeding up a variety of statistical methods involving Monte Carlo, including MCMC \parencite{owen2005quasi}, density estimation \parencite{abdellah2018density}, and data reduction \parencite{mak2018sp}. However, to our knowledge, there has been little-to-no work on integrating QMC ideas to speed up PMC - this is the aim of the current paper. \par

For the proposals, elliptical distributions (e.g., the multivariate normal distribution) are commonly used. Most of the PMC literature focuses on the adaptation of the location (center) parameter, and treats the covariance parameter as static throughout the algorithm. However, the covariance parameter plays a key role in determining the size of the proposal ellipsoid; a poorly chosen covariance could result in a significant mismatch to the target distribution, so the adaptation of this covariance is also essential for the success of PMC. By taking advantage of ISP resampling, we propose a computationally efficient adaptation scheme called \textit{lookback adaptation}. Moreover, since there is adaptation, the samples simulated from the first few iterations are not as good as the later samples. One way to address this is via the weighted PMC estimator \parencite{douc2007dmis2,portier2018ais}, which aims to ``forget" samples from early stages. We further propose a new weighting scheme for the PQMC estimator, which is free of the integrand and the normalizing constant of the target distribution. \par


The paper is organized as follows. Section~\ref{sec:importance_support_points} first reviews QMC and support points, and then introduces the proposed importance support points (ISP). Section~\ref{sec:population_quasi_monte_carlo} discusses the novel Population Quasi-Monte Carlo framework, which makes use of the proposed ISP for resampling and lookback adaptation. Section~\ref{sec:simulation} presents several simulation studies demonstrating the improvements of PQMC over the existing PMC methods. Section~\ref{sec:simulation_pmc_drilling} illustrates the usefulness of PQMC on friction drilling model calibration application, where the posterior is computationally expensive. We conclude the article with some remarks in Section~\ref{sec:conclusion}.

\section{Importance Support Points}
\label{sec:importance_support_points}

We first provide a brief overview of Quasi-Monte Carlo and then introduce  importance support points, which is an integral part of the proposed PQMC framework.

\subsection{Quasi-Monte Carlo}
\label{subsec:quasi-monte_carlo}

Quasi-Monte Carlo (QMC) is traditionally used for numerical integration of a function $h$ with respect to the $p$-dimensional unit hypercube $[0,1]^p$, that is 
\begin{equation}
  \label{eq:qmc1}
  \int_{[0,1]^{p}}h(x) dx \approx \frac{1}{N}\sum_{n=1}^{N}h(x_n)\; .
\end{equation}
In standard Monte Carlo, the $N$ evaluation points $\{x_n\}_{n=1}^{N}$ are sampled uniformly on $[0,1]^{p}$. It is well known that, by Central Limit Theorem, the integration error converges at a rate of $\mathcal{O}(N^{-1/2})$. QMC aims to improve this rate by carefully choosing a set of well-spread out points that fill the $p$-dimensional hypercube in an even and uniform way. This measure of sample uniformity is typically referred to as a \textit{discrepancy measure} in the QMC literature. One well-known discrepancy measure for sample $\{x_n\}_{n=1}^N$ on $[0,1]^p$ is the star-discrepancy \parencite{niederreiter1992random},
\begin{equation}
  \label{eq:qmc2}
  D^{*}_N(\{x_n\}_{n=1}^{N}) = \sup_{a\in[0,1)^{p}}\bigg|\frac{1}{N}\sum_{n=1}^{N}\mathbbm{1}(x_n \in [0,a)) - \prod_{j=1}^{p}a_j\bigg|\;, \quad \mbox{vol}([0,a)) = \prod_{j=1}^{p}a_j \;.
\end{equation}
The star discrepancy measures the maximum difference between the empirical cumulative distribution of the sample $\{x_n\}_{n=1}^{N}$ and the desired uniform distribution on $[0,1]^p$. A small star discrepancy suggests a more uniform sample on $[0,1]^p$, and vice versa. When $p=1$, $D^{*}_N(\{x_n\}_{n=1}^{N})$ reduces to the well-known Kolmogorov-Smirnov statistic for testing the goodness-of-fit of a sample $\{x_n\}_{n=1}^{N}$ to $\mbox{Uniform}[0,1]$ \parencite{owen2013mc}.

The Koksma-Hlawka inequality connects the integration error from \eqref{eq:qmc1} to the star-discrepancy,
\begin{equation}
  \label{eq:qmc3}
  \bigg|\frac{1}{N}\sum_{n=1}^{N}h(x_n) - \int_{[0,1)^{p}}h(x)dx\bigg| \leq D^{*}_N(\{x_n\}_{n=1}^{N})V_{\text{HK}}(h)\; ,
\end{equation}
where $V_{\text{HK}}(h)$ is the total variation of $h$ in the sense of Hardy and Krause for measuring the roughness of integrand $h$ \parencite{owen2013mc}. Equation \eqref{eq:qmc3} shows that samples which are more uniformly distributed over $[0,1]^p$ (i.e., have lower star-discrepancy) tend to result in smaller integration errors. QMC therefore studies sampling strategies which result in low star-discrepancies, as well as other discrepancy measures for which a similar Koksma-Hlawka-like bound holds. These methods achieve an integration rate of $\mathcal{O}(N^{-1}(\log N)^{p-1})$ \parencite{niederreiter1992random} under smoothness assumptions on $h$, which is faster than the MC rate. Recent developments have focused on \textit{randomized} QMC methods, which provide a randomized low-discrepancy sample, with each sample point marginally distributed as $\mbox{Uniform}[0,1]^p$. Randomized QMC allows for unbiased integral estimates, and provides relief from the curse-of-dimensionality for high-dimensional sampling \parencite{dick2013qmc}. We will later make use of randomized QMC for generating proposal samples within PQMC.


One drawback of traditional QMC methods, however, is that they are mainly developed for sampling from the \textit{uniform} unit hypercube. For the resampling step in PQMC, we wish to generate a representative sample from the \textit{non-uniform} distribution for the weighted samples. We introduce next a method called \textit{importance support points}, which achieves this via an extension of a recent QMC method called support points \parencite{mak2018sp}.

\subsection{Importance Support Points}
\label{subsec:importance_support_points}
Let us first review the support points proposed in \textcite{mak2018sp}, which generates representative samples from a target distribution $F$.

\begin{definition}{\textbf{\parencite[Support Points;][]{mak2018sp}}}
\label{df:support_points}
Let $Y\sim F$ where $F$ is a target distribution function on $\emptyset\neq\mathcal{X}\subseteq\mathbb{R}^{p}$ with finite means. The support points $\{\xi_i\}_{i=1}^{n}$ of $F$ are
\begin{equation}
  \label{eq:sp1}
  \{\xi_{i}\}_{i=1}^{n}\in \arg\min_{x_1,\ldots,x_n\in\mathcal{X}}\mathcal{E}(F,F_n) = \arg\min_{x_1,\ldots,x_n\in\mathcal{X}}\frac{2}{n}\sum_{i=1}^{n}\mathbb{E}\lVert x_i - Y\rVert_{2} - \frac{1}{n^2}\sum_{i=1}^{n}\sum_{j=1}^{n}\lVert x_i - x_j\rVert_{2} \; .
\end{equation}  
where $\mathcal{E}(F,F_n)$ is the energy distance \parencite{szekely2004gof,szekely2013es} between $F$ and $F_n$, and $F_n$ is the empirical distribution function for $\{x_i\}_{i=1}^{n}$.
\end{definition}

In the case where only samples $\{y_m\}_{m=1}^{M}$ are available on $F$ (where $M > n$), the Monte Carlo approximation of \eqref{eq:sp1} becomes:
\begin{equation}
  \label{eq:sp2}
  \begin{aligned}
  \{\xi_i\}_{i=1}^{n}= \arg\min_{x_1,\ldots,x_n\in\mathcal{X}}\frac{2}{nM}\sum_{i=1}^{n}\sum_{m=1}^{M}\lVert x_i - y_m\rVert_{2} - \frac{1}{n^2}\sum_{i=1}^{n}\sum_{j=1}^{n}\lVert x_i - x_j\rVert_{2} \; .
  \end{aligned}
\end{equation}
\noindent The first term in \eqref{eq:sp2} forces the support points $\{\xi_{i}\}_{i=1}^{n}$ to mimic the samples from $F$, while the second term forces these points to be as far apart from each other as possible. The latter is often referred to as the ``space-filling property'' in experimental design \parencite{johnson1990design}. The problem in \eqref{eq:sp2} can be efficiently solved via the convex-concave procedure \parencite{yuille2002ccp}, and is implemented in the R package \texttt{support} \parencite{mak2019spR}. 


We now present an extension of support points, called importance support points, which generates representative samples from a \textit{weighted} distribution for $F$. To foreshadow, these ISPs will be used for finding an ``optimal'' subsample from the weighted proposal samples. 

\begin{definition}{\textbf{(Importance Support Points)}}
\label{df:importance_support_points}
Let $\pi = \gamma/Z$ be the probability density function of the target distribution $F$ that we only know up to an unknown constant of proportionality. Let $Y\sim q$, the importance distribution that is defined on the same support $\mathcal{X}$ of $F$. The importance support points $\{\xi_i\}_{i=1}^{n}$ of $F$ with respect to the importance distribution $q$ are
\begin{equation}
  \label{eq:isp1}
  \{\xi_{i}\}_{i=1}^{n} \in \arg\min_{x_1,\ldots,x_n\in\mathcal{X}}\frac{2}{n}\sum_{i=1}^{n}\frac{\mathbb{E}_{q}[w(Y)\lVert x_i - Y\rVert_{2}]}{\mathbb{E}_{q}[w(Y)]} - \frac{1}{n^2}\sum_{i=1}^{n}\sum_{j=1}^{n}\lVert x_i - x_j\rVert_{2}\; ,
\end{equation}
where $w(\cdot) = \gamma(\cdot) / q(\cdot)$ is the unnormalized importance weight function.
\end{definition}


In the case where only samples $\{y_m\}_{m=1}^{M}$ are available from $q$ (where $M > n$), the self-normalized IS approximation of \eqref{eq:isp1} is 
\begin{equation}
  \label{eq:isp2}
  \{\xi_{i}\}_{i=1}^{n} \in \arg\min_{x_1,\ldots,x_n\in\mathcal{X}}\frac{2}{n}\sum_{i=1}^{n}\sum_{m=1}^{M}\bar{w}_m\lVert x_i - y_m\rVert_{2} - \frac{1}{n^2}\sum_{i=1}^{n}\sum_{j=1}^{n}\lVert x_i - x_j\rVert_{2} \; ,
\end{equation}
where $\bar{w}_m = [\gamma(y_m)/q(y_m)]/[\sum_{l=1}^{M}\gamma(y_l)/q(y_l)]$ is the normalized importance weight. This approach only requires samples $\{y_m\}_{m=1}^{M} \sim q$ where $q$ can be some simple distribution that we can generate QMC samples from. The ISP can be generalized to reduce any large set of weighted samples to a few unweighted representative points. The problem in \eqref{eq:isp2} can also be solved via the convex-concave procedure \parencite{yuille2002ccp}, and the details are presented in Appendix~\ref{appendix:importance_support_points}.

\begin{figure}[t!]
  \centering
  
  \begin{subfigure}{0.3\textwidth}
    \centering
    \includegraphics[width=0.9\textwidth]{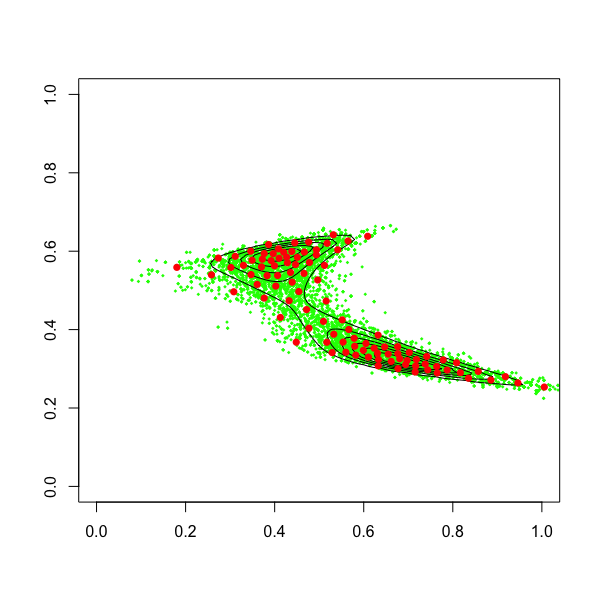}
    \caption{Axe; MCMC}
  \end{subfigure}%
  \begin{subfigure}{0.3\textwidth}
    \centering
    \includegraphics[width=0.9\textwidth]{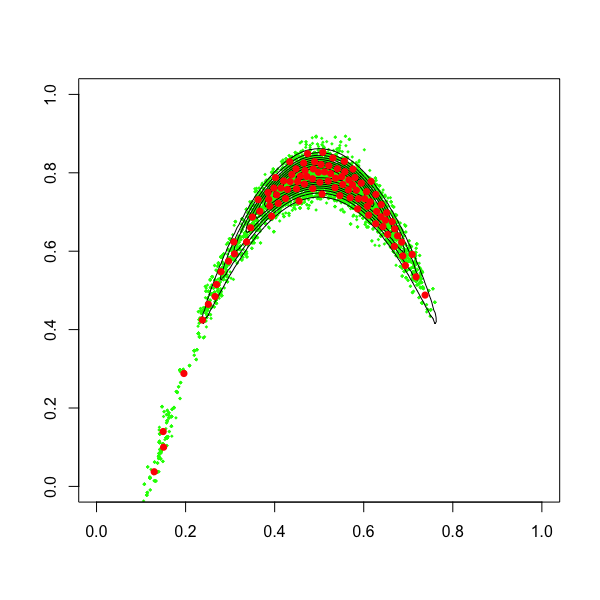}
    \caption{Banana; MCMC}
  \end{subfigure}%
  \begin{subfigure}{0.3\textwidth}
    \centering
    \includegraphics[width=0.9\textwidth]{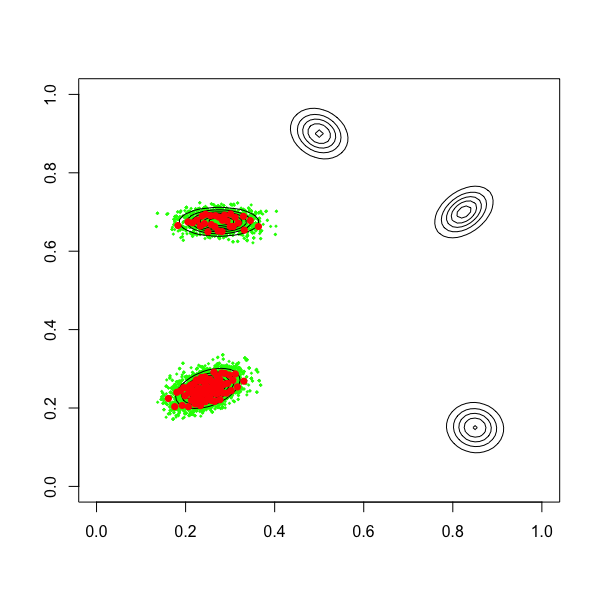}
    \caption{Mixtures; MCMC}
  \end{subfigure}%
  
  \begin{subfigure}{0.3\textwidth}
    \centering
    \includegraphics[width=0.9\textwidth]{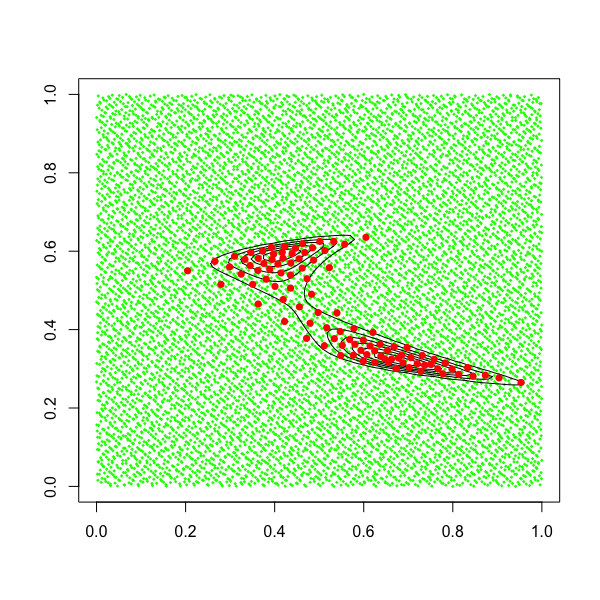}
    \caption{Axe; IS}
  \end{subfigure}%
  \begin{subfigure}{0.3\textwidth}
    \centering
    \includegraphics[width=0.9\textwidth]{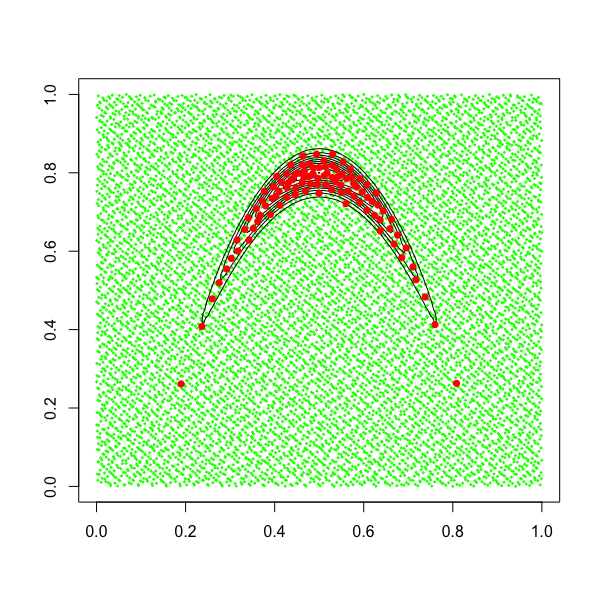}
    \caption{Banana; IS}
  \end{subfigure}%
  \begin{subfigure}{0.3\textwidth}
    \centering
    \includegraphics[width=0.9\textwidth]{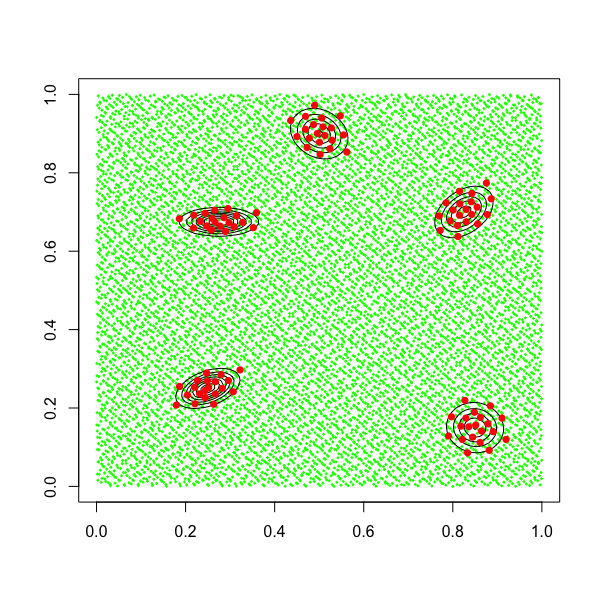}
    \caption{Mixtures; IS}
  \end{subfigure}%
  
  \caption{$n = 100$ support points (red dots) generated from 10{,}000 MCMC samples (green diamonds) and 10{,}000 Importance samples (green diamonds) for two dimensional axe-shaped, banana-shaped, and mixtures of normal distributions. Lines represent the density contours.}
  \label{fig:isp_2d}
\end{figure}

Figure~\ref{fig:isp_2d} shows the $n=100$ support points for the two-dimensional axe-shaped, banana-shaped, and mixtures of normal distributions. Top panels shows the support points from 10{,}000 MCMC samples obtained by running MCMC implemented in the R package \texttt{adaptMCMC} \parencite{scheidegger2018mcmcR} for 15{,}000 iterations and discarding the first 5{,}000 samples as burn-in. Bottom panels shows ISPs from 10{,}000 Sobol' points \parencite{joe2003sobol} generated by R package \texttt{randtoolbox} \parencite{christophe2019randtoolbox} as the importance samples ($q = \mbox{Uniform}[0,1]^2$). When the MCMC explores the distribution well as in the axe-shaped distribution, the support points from MCMC samples are as good as the ISPs. However, for the banana-shaped distribution, the support points from MCMC samples cannot reflect its symmetry structure. The problem is more severe for the mixture of normals as poor mixing on multimodal distribution is a known issue of standard MCMC. ISPs show substantial improvement over support points generated by the MCMC samples, by making use of the density information in $F$.\par

\begin{figure}[t!]
  \centering
  
  \begin{subfigure}{0.3\textwidth}
    \centering
    \includegraphics[width=0.9\textwidth]{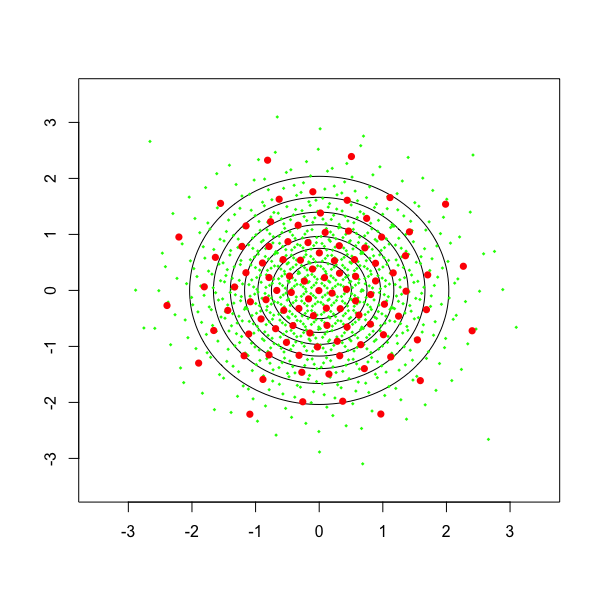}
    \caption{$q = \mathcal{N}(0,I_2)$; ESS = 1{,}000}
  \end{subfigure}%
  \begin{subfigure}{0.3\textwidth}
    \centering
    \includegraphics[width=0.9\textwidth]{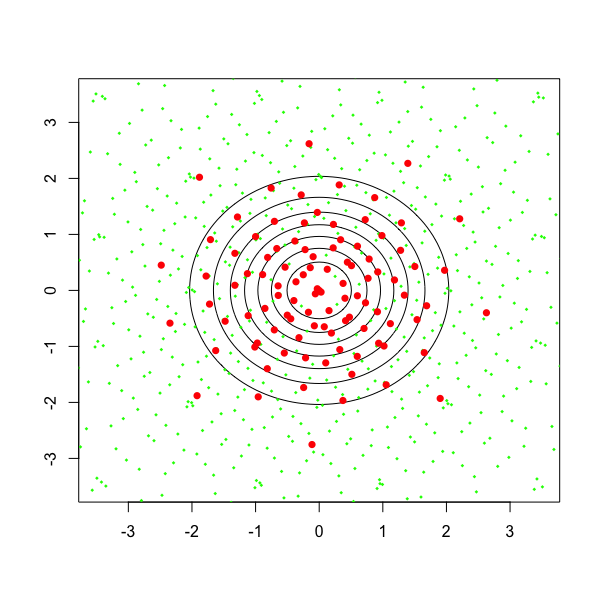}
    \caption{$q = \mathcal{N}(0,3I_2)$; ESS = 209}
  \end{subfigure}%
  \begin{subfigure}{0.3\textwidth}
    \centering
    \includegraphics[width=0.9\textwidth]{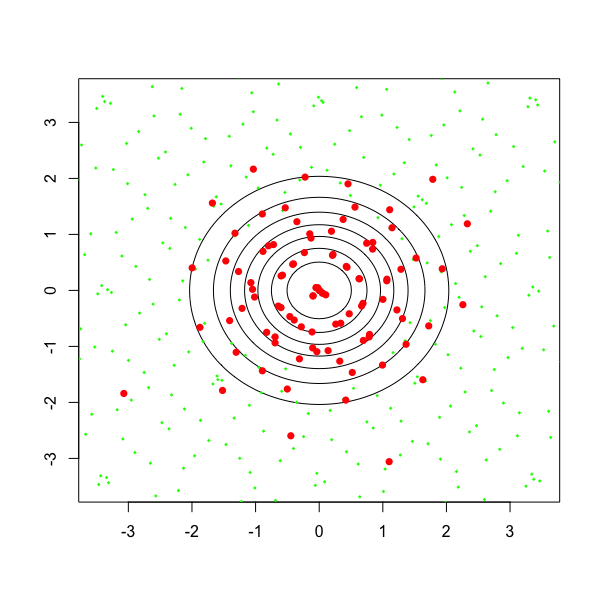}
    \caption{$q = \mathcal{N}(0,5I_2)$; ESS = 78}
  \end{subfigure}%
  
  \caption{$n = 100$ importance support points (red dots) for two dimensional standard normal by 1{,}000 inverse Sobol' points of $q$ as the importance samples (green diamonds). Lines represent the density contours. ESS stands for effective sample size.}
  \label{fig:isp_2d_ess}
\end{figure}

On the other hand, the ISPs suffers the same limitation of IS. The choice of the importance distribution $q$ is critical. As shown in Figure~\ref{fig:isp_2d}, a robust choice would be the uniform distribution over a region that covers the support of the target $\pi$, but we also need sufficient samples on the high-probability regions for yielding good ISPs, where the effective sample size ($N_e = [\sum_{m=1}^{N}\bar{w}_m^2]^{-1}$) is a good measure. Figure~\ref{fig:isp_2d_ess} shows the 100 ISPs for two-dimensional standard normal obtained using 1{,}000 inverse Sobol' points of different importance distributions as the importance samples. The inverse Sobol’ points are generated by first simulating the 1{,}000 Sobol' points on $[0,1]^2$ and then applying the inverse-transform of the desired distribution on those points. As the variance of the proposal increases, fewer importance samples are in the key region, so effective sample size drops and the quality of the ISPs get worse. Thus, the quality of ISPs is subject to the effective sample size of the importance samples, which could be treated as another advantage over the MCMC approach since there is no direct quantitative metric to evaluate the quality of the MCMC samples.\par

Finally, we note that a similar form of the ``weighted'' energy criterion \eqref{eq:isp2} was recently used in \textcite{huling2020energy} to balance covariate distributions for causal inference. The key distinction is that the proposed ISPs optimize for the representative \textit{samples} given fixed weights, whereas the energy balancing weights in \textcite{huling2020energy} optimize for the \textit{weights} given fixed samples. The former problem can be challenging to solve and requires several approximations if we restrict the representative samples $\{\xi_i\}_{i=1}^{n}$ to be points from $\{y_m\}_{m=1}^{M}$ in resampling, which we discuss in the following section. \par

\section{Population Quasi-Monte Carlo}
\label{sec:population_quasi_monte_carlo}

\begin{algorithm}[t!]
\SetAlgoLined
  \textbf{Target Distribution:} $\pi = \gamma / Z$ where $Z$ is the normalizing constant\;
  \textbf{Initialization:} set the parameters for the $K$ initial proposals $\{q_{k}^{(1)}=\mathcal{N}(\cdot|\mu_{k}^{(1)},\Sigma)\}_{k=1}^{K}$ \;
  \vspace{2mm}
  \For{$t = 1,\ldots,T$}{
    $\bullet$ \textbf{Sampling:} draw $J$ samples from each proposal,
    \begin{equation}
      \label{eq:pmc1}
      x_{k,j}^{(t)} \sim q_{k}^{(t)}(x|\mu_{k}^{(t)}, \Sigma)
    \end{equation}
    for $k = 1,\ldots,K$ and $j = 1,\ldots,J$, so total $KJ$ samples are simulated. \\
    $\bullet$ \textbf{Weighting:} compute the importance weight,
    \begin{equation}
      \label{eq:pmc2}
      w_{k,j}^{(t)} = \frac{\gamma(x_{k,j}^{(t)})}{K^{-1}\sum_{i=1}^{K}q_{i}^{(t)}(x_{k,j}^{(t)}|\mu_i^{(t)}, \Sigma)}
    \end{equation}
    for $k = 1,\ldots,K$ and $j = 1,\ldots,J$, and normalize them by 
    \begin{equation}
      \label{eq:pmc3}
      \bar{w}_{k,j}^{(t)} = \frac{w_{k,j}^{(t)}}{\sum_{i=1}^{K}\sum_{l=1}^{J}w_{i,l}}
    \end{equation}
    $\bullet$ \textbf{Adaptation:} perform resampling by drawing $K$ independent samples from the discrete probability random measure
    \begin{equation}
      \label{eq:pmc4}
      \sum_{k=1}^{K}\sum_{j=1}^{J}\bar{w}_{k,j}^{(t)}\delta(x - x_{k,j}^{(t)})
    \end{equation}
    to be the proposal centers $\{\mu_k^{(t+1)}\}_{k=1}^{K}$ for the next iteration.
  }
  \vspace{2mm}
  \textbf{Return:} $\{(x_{k,j}^{(t)}, w_{k,j}^{(t)})\}_{t=1}^{T}{}_{k=1}^{K}{}_{j=1}^{J}$ where $w_{k,j}^{(t)}$ is the unnormalized weight for sample $x_{k,j}^{(t)}$.
 \caption{PMC Algorithm with Normal Proposals and Static Global Covariance}
 \label{algo:pmc}
\end{algorithm}

We now integrate the aforementioned QMC ideas within the PMC framework. For reference, Algorithm~\ref{algo:pmc} outlines the generic PMC procedure, with normal proposal distributions, static global covariance (all proposals share the same covariance), and the deterministic mixture weighting strategy \eqref{eq:pmc2} from \textcite{elvira2017dmmis}. We introduce next novel modifications to incorporate QMC into the \textit{sampling} and \textit{adaptation} steps of PMC, yielding the proposed Population Quasi-Monte Carlo (PQMC) framework. We also consider the use of normal proposals in this paper for ease of illustration, but the results presented can be generalized to any elliptical distribution, such as the multivariate $t$-distribution. Algorithm~\ref{algo:pqmc} outlines the steps for PQMC. We discuss in detail below three novel developments of this PQMC framework: the Quasi-Monte Carlo proposals in the \textit{sampling} step, the importance support point resampling and the lookback adaptation in the \textit{adaptation} step.

\begin{algorithm}[t!]
\SetAlgoLined
  \textbf{Target Distribution:} $\pi = \gamma / Z$ where $Z$ is the normalizing constant\;
  \textbf{Initialization:} set the parameters for the $K$ initial proposals $\{q_{k}^{(1)}=\mathcal{N}(\cdot|\mu_{k}^{(1)},\Sigma^{(1)})\}_{k=1}^{K}$ \;
  \vspace{2mm}
  \For{$t = 1,\ldots,T$}{
    $\bullet$ \textbf{Sampling:} simulate $J$ scrambled Sobol' points $\{u_{k,j}^{(t)}\}_{j=1}^{J}$ and apply equation \eqref{eq:qmcs} to obtain $\{x_{k,j}^{(t)}\}_{j=1}^{n}$ for $k = 1,\ldots,K$, so total $KJ$ samples are simulated. \\
    $\bullet$ \textbf{Weighting:} apply the deterministic mixture weighting strategy as in PMC \eqref{eq:pmc2} and normalize the weights by \eqref{eq:pmc3}. \\
    $\bullet$ \textbf{Adaptation:} perform ISP resampling (Algorithm~\ref{algo:isprs}) with respect to the weighted samples $\{(x_{k,j}^{(t)},\bar{w}_{k,j}^{(t)})\}_{k=1}^{K}{}_{j=1}^{J}$ to obtain new proposal centers $\{\mu_{k}^{(t+1)}\}_{k=1}^{K}$, and apply lookback adaptation \eqref{eq:ca4} for updating the global covariance $\Sigma^{(t+1)}$.
  }
  \vspace{2mm}
  \textbf{Return:} $\{(x_{k,j}^{(t)}, w_{k,j}^{(t)})\}_{t=1}^{T}{}_{k=1}^{K}{}_{j=1}^{J}$ where $w_{k,j}^{(t)}$ is the unnormalized weight for sample $x_{k,j}^{(t)}$.
 \caption{PQMC with Normal Proposals and Adaptive Global Covariance}
 \label{algo:pqmc}
\end{algorithm}

\subsection{Quasi-Monte Carlo Proposals}
\label{subsec:quasi_monte_calo_sampling}
By applying the reparameterization trick, we can represent a $p$-dimensional random variable $x_n\sim\mathcal{N}(\mu,\Sigma)$ by a continuous function defined on a new variable $u_n \sim \mbox{Uniform}[0,1]^{p}$, 
\begin{equation}
  \label{eq:qmcs}
  x_n = g(u_n) = \mu + \Sigma^{1/2}[\Phi^{-1}(u_{n1}),\ldots,\Phi^{-1}(u_{np})]^{T}\; ,
\end{equation}
where $g$ is the inverse transform of the multivariate normal distribution. \textcite{fang1994ntm} show that by applying $g$ to a set of low discrepancy points $\{u_n\}_{n=1}^{N}$, the resulting set $\{x_n = g(u_n)\}_{n=1}^{N}$ also have low F-discrepancy, the largest discrepancy between the cumulative distribution function $F$ and the empirical distribution function $F_{N}$ constructed by $\{x_n\}_{n=1}^{N}$ over the support $\mathcal{X}$. Moreover, from previous discussion of QMC in Subsection~\ref{subsec:quasi-monte_carlo}, randomized QMC is preferred over QMC. Thus, for the sampling step, we use Owen-style scrambling \parencite{owen1998sobol} Sobol' points for $\{u_n\}_{n=1}^{N}$ that is available in the R package \texttt{randtoolbox} \parencite{christophe2019randtoolbox}, and apply \eqref{eq:qmcs} to obtain the samples $\{x_n\}_{n=1}^{N}$. 

\subsection{Importance Support Points Resampling}
\label{subsec:isp_resampling}

Resampling is commonly used for adapting location (center) parameter of the proposals in PMC. It is first introduced by \textcite{rubin1987sir} as Sampling-Importance Resampling, and it plays a key role in Sequential Monte Carlo (SMC) to deal with the weight degeneracy problem \parencite{chen2003smc,del2006smc,cappe2007smc}.

Let $\{\xi_i\}_{i=1}^{n}$ be the resamples for any normalized weighted samples $\{(y_m,\bar{w}_m)\}_{m=1}^{M}$ where $\xi_i \in \{y_m\}_{m=1}^{M}$. Assume that there are $n_m$ copies of $y_m$ in the resamples, i.e., $\sum_{i=1}^{n}\mathbbm{1}(\xi_i = y_m) = n_m$, so $n = \sum_{m=1}^{M} n_m$. The goal is to have the resampled empirical distribution function $\tilde{F}_n(y) = n^{-1}\sum_{i=1}^{n}\mathbbm{1}(\xi_i\leq y) = n^{-1}\sum_{m=1}^{M}n_m\mathbbm{1}(y_m\leq y)$ be as close to the original empirical distribution function $\hat{F}_M(y) = \sum_{m=1}^{M}\bar{w}_m\mathbbm{1}(y_m\leq y)$ as possible. As mentioned in \textcite{hol2006rs}, when the resampled density and the original weighted density are close, we expect that for any integrand $h$, the squared integration error,
\begin{equation}
  \label{eq:isprs1}
  \mathbb{E}\bigg[\bigg(\int h(y)d\tilde{F}_n(y) - \int h(y)d\hat{F}_M(y)\bigg)^2\bigg] = \mathbb{E}\bigg[\bigg(\sum_{m=1}^{M}\frac{n_m - n \bar{w}_m}{n}h(y_m)\bigg)^2\bigg]
\end{equation}
should also be small. In the case of normal proposals, assuming that all covariances are the same, i.e., $\Sigma_{k} = \Sigma$, and considering that $h(\cdot) = \mathcal{N}(x|\cdot,\Sigma)$ for any $x \in \mathcal{X}$, adapting the proposal centers by resampling is to find a mixture of $K$ equally weighted normals, that best approximates the mixture of $KJ$ weighted normals with the centers being the $KJ$ simulated samples and the associated weights computed by \eqref{eq:pmc3}, as the proposal for the next iteration. \textcite{mak2018sp} present a Koksma-Hlawka-like bound that upper bounds the squared integration error \eqref{eq:isprs1} by a term proportional to the energy distance for a large class of integrand $h$. Thus, we propose a deterministic resampling method that find the resampled point set $\{\xi_i\}_{i=1}^{n}$ minimized over the energy distance to the weighted samples $\{(y_m,\bar{w}_m)\}_{m=1}^{M}$, leading to the optimization,
\begin{equation}
  \label{eq:isprs3}
  \begin{aligned}
    \{\xi_{i}\}_{i=1}^{n} \in & \arg\min_{x_1,\ldots,x_n\in\{y_m\}_{m=1}^{M}}\hat{\mathcal{E}}\bigg(\{(y_m,\bar{w}_m)\}_{m=1}^{M}, \{x_i\}_{i=1}^{n}\bigg) \\
    = & \arg\min_{x_1,\ldots,x_n\in\{y_m\}_{m=1}^{M}}\frac{2}{n}\sum_{i=1}^{n}\sum_{m=1}^{M}\bar{w}_m\lVert x_i - y_m\rVert_{2} - \frac{1}{n^2}\sum_{i=1}^{n}\sum_{j=1}^{n}\lVert x_i - x_j\rVert_{2} \; .
  \end{aligned}
\end{equation}
Let us call this the ISP resampling. \eqref{eq:isprs3} is the same optimization problem of the ISPs \eqref{eq:isp2} but under the constraints that $\xi_i \in \{y_m\}_{m=1}^{M}\; \forall i$, making it an integer programming problem that is much harder to solve. \par

\begin{algorithm}[t!]
\SetAlgoLined
  \textbf{Objective:} optimize the ISP problem \eqref{eq:isprs3}. \\
  \vspace{2mm}
  \textbf{Distance Computing:} compute and store the pairwise distances of $\{y_m\}_{m=1}^{M}$. \\
  \vspace{2mm}
  \textbf{Greedy Initialization:} conditional on finding $\{\xi_{j}\}_{j=1}^{i-1}$, $\xi_{i}$ is obtained by
  \begin{equation}
    \label{eq:isprs4}
      \xi_i = \arg\min_{x\in\{y_m\}_{m=1}^{M}}\frac{2}{i}\sum_{m=1}^{M}\bar{w}_m\lVert x - y_m\rVert_{2} - \frac{2}{i^2}\sum_{j=1}^{i-1}\lVert x - \xi_j\rVert_{2}\; .
  \end{equation}
  Solve \eqref{eq:isprs4} for $i=1,\ldots,n$ sequentially to obtain the initial resamples $\{\xi_i\}_{i=1}^{n}$. \\ 
  \vspace{2mm}
  \textbf{Point Refinement:} for $i = 1,\ldots,n$, fixing $\{\xi_{j}\}_{j\neq i}$, refine $\xi_i$ by
  \begin{equation}
    \label{eq:isprs5}
      \xi_i = \arg\min_{x\in\{y_m\}_{m=1}^{M}}\frac{2}{n}\sum_{m=1}^{M}\bar{w}_m\lVert x - y_m\rVert_{2} - \frac{2}{n^2}\sum_{\substack{j=1\\j\neq i}}^{n}\lVert x - \xi_j\rVert_{2}\; .
  \end{equation}
  Repeat above until the energy distance \eqref{eq:isprs3} converges. \\ 
  \vspace{2mm}
  \textbf{Return:} $\{\xi_i\}_{i=1}^{n}$, the set of ISP resamples.
 \caption{Importance Support Points Resampling}
 \label{algo:isprs}
\end{algorithm}

We propose a quadratic runtime sequential optimization procedure presented in Algorithm~\ref{algo:isprs} to approximately solve \eqref{eq:isprs3}. The algorithm consists of three parts: \textit{distance computing}, \textit{greedy initialization}, and \textit{point refinement}. In \textit{distance computing}, we compute and store the pairwise distances of the $M$ samples that are used extensively in the other two parts. We then obtain an initial set of resamples $\{\xi_i\}_{i=1}^{n}$ from \textit{greedy initialization} by sequentially solving \eqref{eq:isprs4} for $i=1,\ldots,n$. \eqref{eq:isprs4} can be solved by first computing the objective value for each $x\in\{y_m\}_{m=1}^{M}$ using pre-computed pairwise distances, and then locating the $y_m$ with smallest objective value. The key idea of the \textit{greedy initialization} is that conditional on having the best $(i-1)$-point resamples $\{\xi_{j}\}_{j=1}^{i-1}$, we find the best $i$-th resample $\xi_{i}$ from $\{y_m\}_{m=1}^{M}$ such that the energy distance between $\{\xi_j\}_{j=1}^{i-1}\cup\{\xi_{i}\}$ and $\{(y_m,\bar{w}_m)\}_{m=1}^{M}$ is minimized. However, the resamples from the \textit{greedy initialization} could be a local optimum. Thus, we propose the \textit{point refinement} to improve each $\xi_i$ by \eqref{eq:isprs5}, that is by fixing the other $(n-1)$ resamples $\{\xi_j\}_{j\neq i}$, we update $\xi_i$ to improve the energy distance \eqref{eq:isprs3}. \eqref{eq:isprs5} is solved similarly by the aforementioned procedure of solving \eqref{eq:isprs4}. In practice, less than 10 repetitions of the \textit{point refinement} step is needed for convergence. The proposed algorithms splits the $n$ variables optimization problem to $n$ smaller optimization problems each with only one variable, making it feasible to solve in polynomial time. The main computational complexity of the algorithm is $\mathcal{O}(M^2)$ from computing the pairwise distance of the $M$ weighted samples. 

The use of the energy distance as the resampling criterion also has a natural connection to the Cram{\'e}r-von Mises criterion \parencite{cramer1928l2d,anderson1962l2d}, a well-known goodness-of-fit measure. Indeed, one can show that the energy distance is a multivariate extension of the Cram{\'e}r-von Mises criterion which preserves rotation-invariance \parencite{szekely2013es}. This new resampling criterion is favored over the traditional approaches for its direct connection to the Cram{\'e}r-von Mises criterion and the minimization of the squared integration error \eqref{eq:isprs1} via Koksma-Hlawka-like bound, whereas the multinomial resampling \parencite{gordon1993mr}, stratified resampling \parencite{kitagawa1996sr}, residual resampling \parencite{liu1998rr}, and systematic resampling \parencite{carpenter1999sr}, all aim to minimize
\begin{equation}
  \label{eq:isprs2}
  \mathbb{E}[(n_m - n\bar{w}_m)^2] = (\mathbb{E}[n_m] - n\bar{w}_m)^2 + \mathbb{V}[n_m] = \mathbb{V}[n_m]
\end{equation}
where $\mathbb{E}[n_m] = n\bar{w}_m$ unbiased \parencite{douc2005rs}. Though ISP resampling requires quadratic runtime, in PMC, the number of samples simulated at each iteration, $M = KJ$, is moderate size, making the use of quadratic runtime algorithm acceptable. Also, with the ISP resampling, we can allow $K$ to be small without losing too much information in the resampling step, so the additional computational burden can be offset by the reduction in computational cost from the $\mathcal{O}(K^2J)$ evaluations of the proposal distributions in the deterministic mixture weighting strategy. \par

\begin{figure}[t!]
  \centering
  
  \begin{subfigure}{0.3\textwidth}
    \centering
    \includegraphics[width=0.9\textwidth]{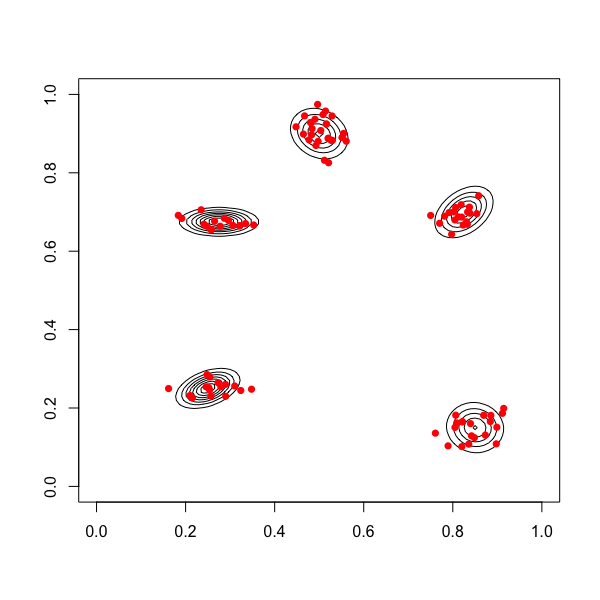}
    \caption{Multinomial}
  \end{subfigure}%
  \begin{subfigure}{0.3\textwidth}
    \centering
    \includegraphics[width=0.9\textwidth]{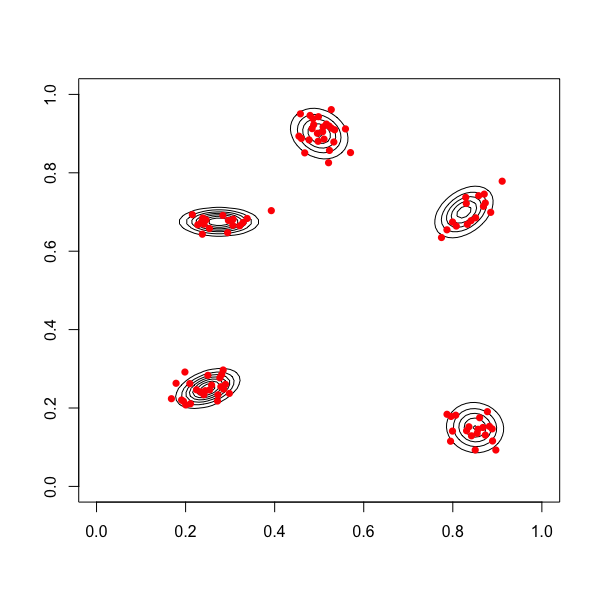}
    \caption{Systematic}
  \end{subfigure}%
  \begin{subfigure}{0.3\textwidth}
    \centering
    \includegraphics[width=0.9\textwidth]{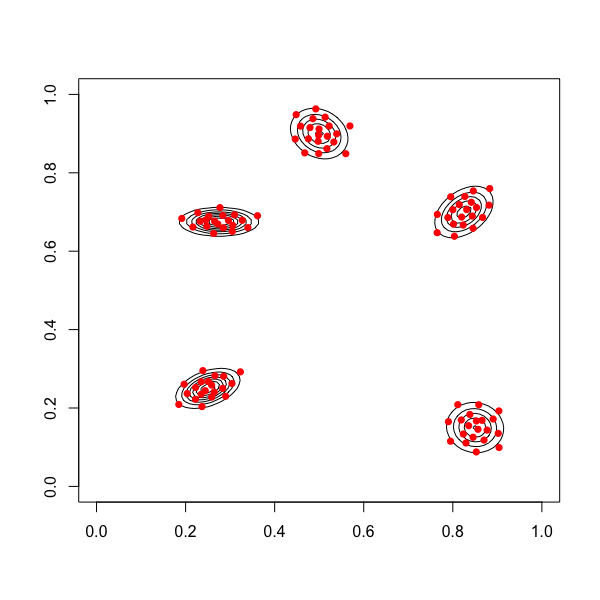}
    \caption{ISP}
  \end{subfigure}%
  
  \caption{$n = 100$ resamples from 10{,}000 Sobol' points as importance samples for the mixture of normals using multinomial, systematic, and ISP. Lines represent the density contours.}
  \label{fig:resample_2d}
\end{figure}

Figure~\ref{fig:resample_2d} shows the 100-point resampled point set for the mixture of normals from the importance samples of 10{,}000 Sobol' points over $[0,1]^2$ using multinomial, systematic, and ISP resampling. By visualization, the 100 points from ISP resampling serve as a better set of the proposal centers, since these points not only better capture the shape of the target distribution, but are also well-spaced out from one another (``space-filling''), which allows for better exploration.\par

\subsection{Covariance Adaptation}
\label{subsec:covariance_adaptation}
Finally, we present the adaptation procedures for updating covariance matrices in the proposal distributions, which is critical for the success of PQMC. At each iteration, the set of equally weighted proposals can be seen as a kernel density approximation of the target distribution where each proposal plays the role of kernel \parencite{elvira2017dmmis}. However, finding the optimal kernel covariances often relies on cross-validation, which is computational expensive. On the other hand, as proposed by \textcite{cappe2008mmis} for Mixture PMC and by \textcite{ji2013amcmc} for adaptive MCMC, an alternative solution is to find the set of covariances $\{\Sigma_{k}^{(t+1)}\}_{k=1}^{K}$ that minimizes the Kullback-Leibler (KL) divergence between the target density $\pi$ and the normal mixture proposal $K^{-1}\sum_{k=1}^{K}\mathcal{N}(x|\mu_{k}^{(t+1)},\Sigma_{k}^{(t+1)})$ for next iteration if the proposals have different covariances,
\begin{equation}
  \label{eq:ca1}
  \begin{aligned}
    \{\Sigma_k^{(t+1)}\}_{k=1}^{K} \in& \arg\min_{C_{1},\ldots,C_{K}\in S^{p}_{+}} KL\bigg(\pi(x) \bigg|\bigg| \frac{1}{K}\sum_{k=1}^{K}\mathcal{N}(x|\mu_{k}^{(t+1)},C_{k})\bigg) \\
    =& \arg\min_{C_{1},\ldots,C_{K}\in S^{p}_{+}}\bigg(\int_{\mathcal{X}}\pi(x)\log\pi(x)dx - \int_{\mathcal{X}}\pi(x)\log\bigg[\frac{1}{K}\sum_{k=1}^{K}\mathcal{N}(x|\mu_{k}^{(t+1)},C_{k})\bigg] dx\bigg) \\
    =& \arg\max_{C_{1},\ldots,C_{K}\in S^{p}_{+}} \int_{\mathcal{X}}\pi(x)\log\bigg[\frac{1}{K}\sum_{k=1}^{K}\mathcal{N}(x|\mu_{k}^{(t+1)},C_{k})\bigg] dx\; ,
  \end{aligned}
\end{equation}
where $\{\mu_k^{(t+1)}\}_{k=1}^{K}$ are obtained from resampling. Recall that at the $t$-th iteration, we have a set of weighted samples $\{(x_{k,j}^{(t)}, \bar{w}_{k,j}^{(t)})\}_{k=1}^{K}{}_{j=1}^{J}$ that approximately simulated from $\pi$, and thus leading to the Importance Sampling approximation of \eqref{eq:ca1},
\begin{equation}
  \label{eq:ca2}
   \{\Sigma_k^{(t+1)}\}_{k=1}^{K} \in \arg\min_{C_{1},\ldots,C_{K}\in S^{p}_{+}}\sum_{k=1}^{K}\sum_{j=1}^{J}\bar{w}_{k,j}^{(t)}\log\bigg[\frac{1}{K}\sum_{i=1}^{K}\mathcal{N}(x_{k,j}^{(t)}|\mu_{i}^{(t+1)},C_{i})\bigg]\; .
\end{equation}
$\{\Sigma_k^{(t+1)}\}_{k=1}^{K}$ can be estimated by applying Expectation-Maximization \parencite[EM;][]{dempster1977em,wu1983em} to the Gaussian Mixtures Model with fixed weights $1/K$ and fixed centers $\{\mu_k^{(t+1)}\}_{k=1}^{K}$ on the $t$-th iteration's weighted samples $\{(x_{k,j}^{(t)}, \bar{w}_{k,j}^{(t)})\}_{k=1}^{K}{}_{j=1}^{J}$. The EM should converge in around 10 steps since only covariances are estimated. Let us call it exact covariance adaptation. However, this is computational expensive as each EM step requires $\mathcal{O}(K^2J)$ evaluations of the proposal distribution. \par

Consider a special case when all the proposals in the same iteration share one global covariance matrix, i.e., $\Sigma_1^{(t+1)} = \cdots = \Sigma_K^{(t+1)} = \Sigma^{(t+1)}$. We propose the lookback covariance adaptation that does not require additional evaluations of the proposal distribution. The idea is that after several iterations of the PQMC, the samples should converge to the desired regions, then the proposal centers will not vary much from iteration to iteration except in different orientation when ISP resampling is used. Thus, the lookback covariance adaptation optimizes over the prior centers $\{\mu_k^{(t)}\}_{k=1}^{K}$, which gives
\begin{equation}
  \label{eq:ca3}
  \Sigma^{(t+1)} = \arg\min_{C\in S^{p}_{+}}\sum_{k=1}^{K}\sum_{j=1}^{J}\bar{w}_{k,j}^{(t)}\log\bigg[\frac{1}{K}\sum_{i=1}^{K}\mathcal{N}(x_{k,j}^{(t)}|\mu_{i}^{(t)},C)\bigg]\; .
\end{equation}
We do an one-step EM using $\Sigma^{(t)}$ as the prior, leading to a closed-form update,
\begin{equation}
  \label{eq:ca4}
  \Sigma^{(t+1)} = \sum_{k=1}^{K}\sum_{j=1}^{J}\bar{w}_{k,j}^{(t)}\frac{\mathcal{N}(x_{k,j}^{(t)}|\mu_{k}^{(t)},\Sigma_{k}^{(t)})}{\sum_{i=1}^{K}\mathcal{N}(x_{k,j}^{(t)}|\mu_{i}^{(t)},\Sigma_{i}^{(t)})}(x_{k,j}^{(t)} - \mu_{k}^{(t)})(x_{k,j}^{(t)} - \mu_{k}^{(t)})^{T} \; ,
\end{equation}
where the evaluations of the proposal distributions are all done in the weighting steps. Though the lookback covariance adaptation can also work jointly with the traditional resampling methods, the performance would not be as good since it builds on the assumption that the proposal centers are not varying much except in different rotation as the algorithm converge, where the traditional resampling methods might not be able to achieve that due to the lack of consideration on the space-filling property of the resamples. 

\subsection{Weighted PMC Estimator}
\label{subsec:weighted_pmc}

With the above modifications, the PQMC method (Algorithm \ref{algo:pqmc}) returns a set of weighted samples $\{(x_{k,j}^{(t)}, w_{k,j}^{(t)})\}_{t=1}^{T}{}_{k=1}^{K}{}_{j=1}^{J}$. This can then be used to construct the following PQMC estimator for $\mathbb{E}_{\pi}[h(X)]$ for a desired integrand $h$. When the normalizing constant $Z$ is known, the standard PMC estimator for $\mathbb{E}_{\pi}[h(X)]$ is
\begin{equation}
  \label{eq:pmce1}
  \hat{I}^{\text{PMC}} = \frac{1}{Z}\bigg(\frac{1}{TKJ}\sum_{t=1}^{T}\sum_{k=1}^{K}\sum_{j=1}^{J}w_{k,j}^{(t)}h(x_{k,j}^{(t)})\bigg) = \frac{1}{T}\sum_{t=1}^{T}\hat{I}^{\text{PMC}}_{t}\; ,
\end{equation}
where $\hat{I}^{\text{PMC}}_{t} = \frac{1}{Z}(\frac{1}{KJ}\sum_{k=1}^{K}\sum_{j=1}^{J}w_{k,j}^{(t)}h(x_{k,j}^{(t)}))$ is the estimator constructing using only the $t$-th iteration weighted samples. If $Z$ is unknown, we can replace it by a consistent estimator
\begin{equation}
  \label{eq:pmce2}
  \hat{Z}^{\text{PMC}} = \frac{1}{TKJ}\sum_{t=1}^{T}\sum_{k=1}^{K}\sum_{j=1}^{J}w_{k,j}^{(t)}\; .
\end{equation}
We can see that the standard PMC estimator can be viewed as the simple average of $T$ different estimators each is constructed by the weighted samples simulated from the corresponding iteration.

However, when there is adaptation, the standard PMC estimator is not efficient since better samples are obtained as the algorithm proceeds. The weighted PMC (WPMC) estimator assigns a set of correction weights $\{\alpha^{(t)}\}_{t=1}^{T}$ with the constraint that $\sum_{t=1}^{T}\alpha^{(t)} = 1$ to the $T$ estimators, allowing to ``forget" the poor samples simulated at the early stages. When the normalizing constant $Z$ is known, the WPMC estimator for $\mathbb{E}_{\pi}[h(X)]$ is
\begin{equation}
  \label{eq:wpmce1}
  \hat{I}^{\text{WPMC}} = \frac{1}{Z}\sum_{t=1}^{T}\alpha^{(t)}\hat{I}^{\text{PMC}}_{t} = \frac{1}{Z}\bigg(\frac{1}{KJ}\sum_{t=1}^{T}\sum_{k=1}^{K}\sum_{j=1}^{J}\alpha^{(t)}w_{k,j}^{(t)}h(x_{k,j}^{(t)})\bigg)\; .
\end{equation}
If $Z$ is unknown, we replace it by the following consistent estimator
\begin{equation}
  \label{eq:wpmce2}
  \hat{Z}^{\text{WPMC}} = \frac{1}{KJ}\sum_{t=1}^{T}\sum_{k=1}^{K}\sum_{j=1}^{J}\alpha^{(t)}w_{k,j}^{(t)}\; .
\end{equation}
Let $N_e^{(t)}$ denotes the effective sample size of the weighted samples simulated at the $t$-th iteration. We propose the correction weights $\{\alpha^{(t)}\}_{t=1}^{T}$ that approximately minimize the variance of $\hat{I}^{\text{WPMC}}$,
\begin{equation}
  \label{eq:wpmce3}
  \alpha^{(t)} = \frac{N_e^{(t)}}{\sum_{i=1}^{T} N_e^{(i)}}\; .
\end{equation}
The proposed weights are proportional to the effective sample size, assigning larger weights to the estimators that are more reliable. Appendix~\ref{appendix:wpmc} provides further justification of these weights. This approach is free from the integrand $h$ and does not require knowing the normalizing constant. The idea of using effective sample size to weight the estimators from different iterations is also mentioned in the Adaptive Population Importance Sampler \parencite{martino2015apis}.

\section{Simulation Results}
\label{sec:simulation}

In this section, we report some simulation results to demonstrate the improvement of our proposed importance support points resampling and Population Quasi-Monte Carlo algorithm. More simulations results can be found in Appendix~\ref{appendix:simulation}. Source codes and tutorials can be found at \url{https://github.com/BillHuang01/PQMC}.

\subsection{Importance Support Points Resampling}
\label{subsec:simulation_resampling}

\begin{figure}[t!]
  \centering
  \includegraphics[width=0.9\textwidth]{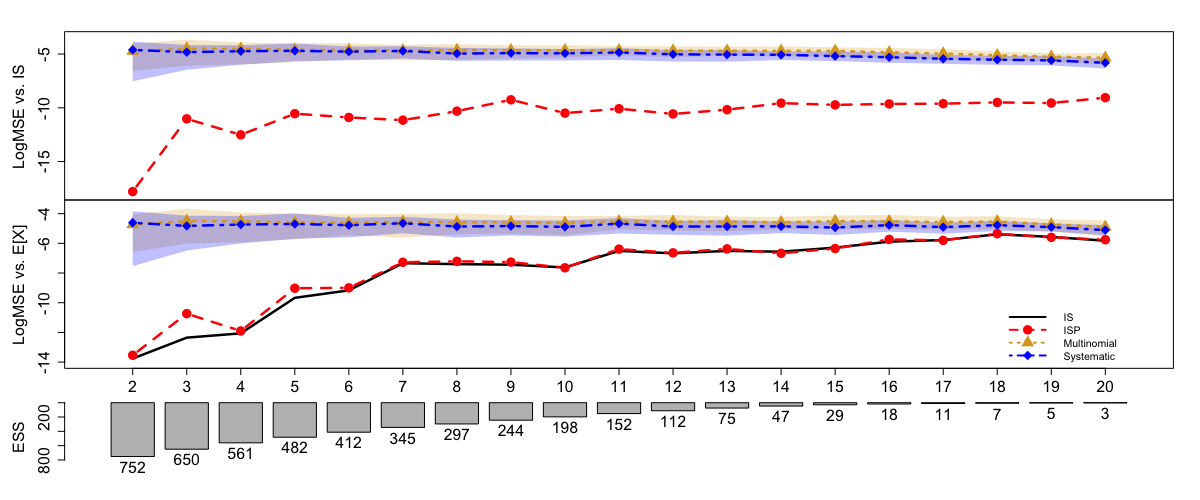}
  \caption{LogMSEs in the estimation of the IS estimator and $\mathbb{E}[X]$ where $X\sim\mathcal{N}(0,I_{p})$ for $p = 2,\ldots,20$ using 100 resampled points from the 1{,}000 inverse Sobol' points of $q = \mathcal{N}(0,\sqrt{2}I_{p})$ as the importance samples by multinomial, systematic, and ISP resampling. MSE for multinomial and systematic are averaged over 100 independent runs. ESS stands for effective sample size. Lines denote the logMSEs, and shaded bands mark the 10th and 90th quantiles.}
  \label{fig:resample_ess}
\end{figure}

Let $X\sim\mathcal{N}(0,I_{p})$, the $p$-dimensional standard normal distribution, be the target distribution. Consider $M = 1{,}000$ inverse Sobol' points of the importance distribution $\mathcal{N}(0,\sqrt{2}I_{p})$ as the importance samples $\{(y_m,\bar{w}_m)\}_{m=1}^{M}$ where $\bar{w}_m$ is the normalized importance weight for $y_m$, then $\hat{I}_{M} = \sum_{m=1}^{M}\bar{w}_m y_m$ is the Importance Sampling (IS) estimator for $\mathbb{E}[X]$. Next, we simulate $n=100$ resamples $\{\xi_i\}_{i=1}^{n}$ using multinomial, systematic, and ISP resampling on the importance samples, then $\tilde{I}_n = n^{-1}\sum_{i=1}^{n}\xi_i$ is the Monte Carlo (MC) estimator for $\mathbb{E}[X]$. We repeat it 100 times to obtain 100 MC estimators $\{\tilde{I}_n^{(l)}\}_{l=1}^{100}$, then $\mbox{MSE}(\hat{I}_N) = 100^{-1}\sum_{l=1}^{100}p^{-1}(\tilde{I}_n^{(l)} - \hat{I}_N)^{T}(\tilde{I}_n^{(l)} - \hat{I}_N)$ is a good empirical approximation for the squared integration error in \eqref{eq:isprs1} with $h(y) = y$ where the error is averaging over $p$ components. The top panel of Figure~\ref{fig:resample_ess} shows the $\mbox{MSE}(\hat{I}_N)$ in log for $p = 2,\ldots,20$. Empirically, we see that ISP resampling enjoys the squared integration error of $\mathcal{O}(n^{-3})$ in 2 dimensions and $\mathcal{O}(n^{-2})$ up to 20 dimensions, outperforming the other two resampling methods. However, the ISP resampling suffers from small effective sample size as dimension increases. One might also be interested in how well the MC estimator using $\{\xi_i\}_{i=1}^{n}$ approximates $\mathbb{E}[X]$. A good empirical measure is $\mbox{MSE}(\mathbb{E}[X]) = 100^{-1}\sum_{l=1}^{100}p^{-1}(\tilde{I}_n^{(l)} - \mathbb{E}[X])^{T}(\tilde{I}_n^{(l)} - \mathbb{E}[X])$. The middle panel of Figure~\ref{fig:resample_ess} shows the $\mbox{MSE}(\mathbb{E}[X])$ in log for $p = 2,\ldots,20$. The performance of the estimator using the 100 resamples from ISP resampling is almost as good as the performance of the IS estimator using the 1{,}000 importance samples, showing that ISP resampling can retain most information from the original importance samples $\{(y_m,\bar{w}_m)\}_{m=1}^{M}$. Consider a more carefully chosen importance distribution $q = \mathcal{N}(0,3^{(2/p^{0.8})}I_{p})$ such that the effective sample sizes are similar for $p = 2,\ldots,20$ (Figure~\ref{fig:resample_rhd} in Appendix~\ref{appendix:simulation}): the ISP resampling only suffers slightly from the curse of dimensionality. Thus, with proper adaptation for the proposals in PQMC, the ISP resampling appears to be quite robust for this high dimensional problem. We only compare the ISP resampling to multinomial resampling for its simplicity and systematic resampling for its good empirical performance mentioned in the literatures \parencite[e.g.][]{douc2005rs}. \par

\subsection{Two Dimensional PQMC Example}
\label{subsec:simulation_pmc_2d}

Consider a two-dimensional multimodal distribution that consists of a mixture of five normals, 
\begin{equation}
  \label{eq:pmc_2d}
  \pi(x) = \frac{1}{5}\sum_{i=1}^{5}\mathcal{N}(x|\mu_i,\Sigma_i)\; ,
\end{equation}
where $\mu_1 = [0.250,0.250]^{T}$, $\mu_2 = [0.500,0.900]^{T}$, $\mu_3 = [0.825,0.700]^{T}$, $\mu_4 = [0.275, 0.675]^{T}$, $\mu_5 = [0.850, 0.150]^{T}$, $\Sigma_1 = 40^{-2}[2,0.6;0.6,1]$, $\Sigma_2 = 40^{-2}[2, -0.4; -0.4,2]$, $\Sigma_3 = 40^{-2}[2,0.8;0.8,2]$, $\Sigma_4 = 40^{-2}[3,0;0,0.5]$, and $\Sigma_5 = 40^{-2}[2,-0.1;-0.1,2]$. The example is from \textcite{elvira2017dmmis} but with proper scaling so the main support of $\pi$ is inside $[0,1]^2$, and the density contour is shown in Figure~\ref{fig:resample_2d}. The mean $\mathbb{E}_{\pi}[X] = [0.540,0.535]^{T}$ and the normalizing constant $Z = 1$ can both be computed analytically so we can validate the performance of the PQMC and PMC. We use the Mean Squared Error (MSE) of the estimates as the evaluation metric. \par

Let us compare the PQMC described in Algorithm~\ref{algo:pqmc} to the generic PMC outlined in Algorithm~\ref{algo:pmc} both with normal proposals and global covariance. For the PMC, we consider two resampling methods: multinomial and systematic. We also apply the lookback covariance adaptation to the PMC. We run both PMC and PQMC for $T = 10$ iterations but vary $K$ and $J$ while keeping $KJ = 1{,}000$, leading to the total of $TKJ = 10{,}000$ evaluations of the target distribution. The initial proposal centers are selected as the $K$ Sobol' points over $[0,1]^2$. We use the same isotropic covariance matrix $\sigma^2 I_2$ for all the initial proposals, i.e. $\Sigma_{k}^{(1)} = \sigma^2 I_2 \; \forall k$, with $\sigma = 0.1,0.2,0.5$. For schemes without covariance adaptation, we fix the covariances for all iterations as in \textcite{elvira2017dmmis}, i.e. $\Sigma_{k}^{(t)} = \sigma^2 I_2 \; \forall k,t$ with the specified $\sigma$. For the ones with lookback adaptation, we keep the adapted covariance isotropic for simplicity, i.e. $\Sigma_{k}^{(t)} = (\sigma^{(t)})^2 I_2 \; \forall k,t$ where the adaptation is performed on $\sigma^{(t)}$ only. We compute the MSEs for both the standard PMC estimator and the weighted PMC estimator averaging over 100 independent runs. \par

\begin{table}[t!]
  \centering
  \resizebox{\columnwidth}{!}{%
  \begin{tabular}{|cccc|ccc|}
  \hline
Estimator & Algorithm & K & J & $\sigma = 0.1$ & $\sigma = 0.2$ & $\sigma = 0.5$ \\
\hline
Standard & PMC (Multinomial) & 25 & 40 & -8.09 [-16.09,-5.07] & -8.56 [-12.98,-6.53] & -8.14 [-14.36,-6.58] \\
Standard & PMC (Systematic) & 25 & 40 & -8.78 [-14.32,-5.74] & -8.71 [-14.14,-7.22] & -8.17 [-13.48,-6.19] \\
Standard & PMC (Multinomial + Lookback) & 25 & 40 & -7.65 [-14.67,-5.38] & -7.78 [-13.79,-5.33] & -8.71 [-13.15,-5.70] \\
Standard & PMC (Systematic + Lookback) & 25 & 40 & -8.30 [-16.08,-5.02] & -8.35 [-13.94,-5.55] & -8.62 [-17.77,-5.80] \\
Standard & PQMC (ISP + Lookback) & 25 & 40 & -12.33 [-15.82,-10.03] & -11.18 [-16.12,-9.71] & -9.63 [-14.32,-7.58] \\
\hdashline
Weighted & PMC (Multinomial) & 25 & 40 & -7.97 [-14.49,-5.02] & -8.51 [-12.58,-6.54] & -8.12 [-12.68,-6.50] \\
Weighted & PMC (Systematic) & 25 & 40 & -8.65 [-14.86,-5.60] & -8.65 [-13.08,-7.07] & -8.13 [-12.95,-6.31] \\
Weighted & PMC (Multinomial + Lookback) & 25 & 40 & -7.44 [-15.62,-5.23] & -7.45 [-14.44,-4.92] & -7.98 [-15.27,-4.98] \\
Weighted & PMC (Systematic + Lookback) & 25 & 40 & -8.05 [-18.16,-4.94] & -7.91 [-16.29,-5.01] & -8.14 [-15.74,-5.06] \\
Weighted & PQMC (ISP + Lookback) & 25 & 40 & \textcolor{red}{\textbf{-15.04}} [-20.62,-13.35] & \textcolor{red}{\textbf{-14.54}} [-18.66,-13.20] & \textcolor{red}{\textbf{-13.81}} [-18.44,-12.02] \\
\hline
Standard & PMC (Multinomial) & 50 & 20 & -9.87 [-14.56,-8.41] & -8.79 [-13.66,-7.22] & -8.02 [-13.36,-6.12] \\
Standard & PMC (Systematic) & 50 & 20 & -10.13 [-13.36,-8.92] & -8.90 [-12.88,-7.27] & -7.95 [-12.39,-6.40] \\
Standard & PMC (Multinomial + Lookback) & 50 & 20 & -10.71 [-14.68,-9.02] & -10.31 [-13.66,-8.48] & -9.03 [-14.87,-5.70] \\
Standard & PMC (Systematic + Lookback) & 50 & 20 & -11.03 [-17.20,-9.43] & -10.14 [-13.85,-8.80] & -9.25 [-16.24,-7.67] \\
Standard & PQMC (ISP + Lookback) & 50 & 20 & -11.98 [-16.74,-10.40] & -11.01 [-15.59,-9.26] & -9.39 [-14.29,-7.50] \\
\hdashline
Weighted & PMC (Multinomial) & 50 & 20 & -9.99 [-14.50,-8.61] & -8.67 [-13.80,-7.01] & -8.01 [-13.23,-6.17] \\
Weighted & PMC (Systematic) & 50 & 20 & -10.15 [-16.04,-8.83] & -8.82 [-12.89,-7.17] & -7.92 [-13.42,-6.50] \\
Weighted & PMC (Multinomial + Lookback) & 50 & 20 & -11.83 [-17.14,-9.66] & -11.93 [-16.13,-10.03] & -9.45 [-16.80,-4.99] \\
Weighted & PMC (Systematic + Lookback) & 50 & 20 & -12.25 [-16.88,-10.86] & -11.85 [-17.57,-10.19] & -11.57 [-15.62,-9.85] \\
Weighted & PQMC (ISP + Lookback) & 50 & 20 & \textbf{-14.89} [-19.18,-13.25] & \textbf{-14.35} [-18.07,-12.61] & \textbf{-13.11} [-17.29,-11.55] \\
\hline
Standard & PMC (Multinomial) & 100 & 10 & -9.78 [-15.37,-7.89] & -9.01 [-13.16,-7.47] & -7.92 [-11.18,-6.49] \\
Standard & PMC (Systematic) & 100 & 10 & -10.12 [-15.29,-8.24] & -8.97 [-12.88,-7.71] & -8.05 [-11.68,-6.49] \\
Standard & PMC (Multinomial + Lookback) & 100 & 10 & -10.99 [-15.21,-8.59] & -10.29 [-15.70,-8.78] & -9.19 [-14.28,-7.23] \\
Standard & PMC (Systematic + Lookback) & 100 & 10 & -10.77 [-17.62,-9.13] & -10.18 [-15.17,-8.41] & -9.29 [-14.83,-7.80] \\
Standard & PQMC (ISP + Lookback) & 100 & 10 & -11.65 [-18.61,-9.59] & -10.84 [-14.62,-9.13] & -9.52 [-14.28,-7.74] \\
\hdashline
Weighted & PMC (Multinomial) & 100 & 10 & -9.82 [-15.40,-7.54] & -8.99 [-13.54,-7.42] & -7.88 [-12.04,-6.50] \\
Weighted & PMC (Systematic) & 100 & 10 & -10.23 [-14.81,-8.35] & -8.92 [-12.98,-7.50] & -7.99 [-11.90,-6.45] \\
Weighted & PMC (Multinomial + Lookback) & 100 & 10 & -12.46 [-15.29,-11.10] & -12.36 [-19.14,-10.70] & -11.80 [-18.31,-10.16] \\
Weighted & PMC (Systematic + Lookback) & 100 & 10 & -12.51 [-16.59,-11.24] & -12.20 [-16.80,-10.66] & -11.76 [-15.13,-9.88] \\
Weighted & PQMC (ISP + Lookback) & 100 & 10 & \textbf{-14.34} [-19.00,-13.08] & \textbf{-13.89} [-19.40,-12.41] & \textbf{-12.89} [-18.01,-11.05] \\
\hline
  \end{tabular}
  }
  \caption{LogMSEs in the estimation of $\mathbb{E}_{\pi}[X]$ for the two dimensional mixture of five normals using different values of $K$, $J$, and $\sigma$ with the initial proposal centers being the $K$ Sobol' points over $[0,1]^2$. The number of evaluations of the target distribution is fixed to $TKJ = 10{,}000$. The MSEs are averaged over 100 independent runs and shown in log under format ``mean [min,max]". The best results for each value of $\sigma$ are highlighted in red bold-face.}
  \label{tab:pmc_2d_full_m}
\end{table}

Table~\ref{tab:pmc_2d_full_m} shows the MSEs in log for the estimation of $\mathbb{E}_{\pi}[X]$. The weighted PMC estimator on PQMC samples outperforms all PMC settings for different values of $K$, $J$, and $\sigma$, demonstrating the significant improvement from the PQMC. Moreover, PQMC is robust even for small $K$. Recall that the deterministic mixture weighting scheme requires $\mathcal{O}(K^2J)$ evaluations of the proposal distributions, thus by being able to use a small $K$, PQMC could reduce the computational cost of proposal evaluations, somewhat offsetting the additional computational burden the ISP resampling brings over the traditional resampling methods. For the PMC algorithm, having the lookback covariance adaptation generally improves the performance, especially when the initial $\sigma$ is chosen poorly. Also, when the $\sigma$ is adapted, weighted PMC estimator is preferred. However, when the number of proposals $K$ is small, using lookback covariance adaptation in PMC could sometimes go wrong for multimodal distribution. The reason is that if too many samples (much larger than $K$) are in the high density regions, random resampling likely results in $K$ particles that are from only few modals rather from all modals in which there exists some samples. The same issue also causes the worse weighted estimator for PMC samples when $K$ is small. Because of its space-filling property, ISP resampling in PQMC does not suffer from the aforementioned issue. Table~\ref{tab:pmc_2d_full_z} in the Appendix shows the MSEs in log for the estimation of the normalizing constant $Z$, and similar conclusions can be drawn. \par

Now consider a ``bad" initialization by using $K$ Sobol' points over $[0.4,0.6]^2$ for the initial proposal centers as in \textcite{elvira2017dmmis} to further test the robustness of PQMC. Table~\ref{tab:pmc_2d_sub_m} in the Appendix shows the MSEs in log for the estimation of $\mathbb{E}_{\pi}[X]$ using the ``bad" initialization. When $\sigma = 0.2 \mbox{ or } 0.5$, the weighted PMC estimator on PQMC sample again outperforms the PMC for different values of settings of $K$ and $J$. When the initial $\sigma = 0.1$ is too small, the performance of both PMC and PQMC are bad since they both fail to discover all the modes of the target distribution. Similar conclusion can be drawn from the MSEs in log for the estimation of the normalizing constant $Z$ presented in Table~\ref{tab:pmc_2d_sub_z} in the Appendix. This shows that the proposed PQMC is robust against the ``bad" initialization of the proposal centers as long as the initial proposal covariances are large enough such that at least few of the simulated samples at the initial iteration can land on the key regions.

\subsection{High Dimensional PQMC Example}
\label{subsec:simulation_pmc_hd}

Consider a ten-dimensional multimodal distribution that consists of a mixture of three normals, 
\begin{equation}
  \label{eq:mixture_10d}
  \pi(x) = \frac{1}{3}\sum_{i=1}^{3}\mathcal{N}(x|\mu_i,\Sigma_i)\; ,
\end{equation}
where $\mu_{1,j} = 0.375$ for $j = 1,\ldots,10$, $\mu_{2,j} = 0.575$ for $j = 1,\ldots,10$, and $\mu_{3,j} = 0.700$ for $j = 1,\ldots,10$, $\Sigma_{1} = \Sigma_{2} = \Sigma_{3} = 0.2^2 I_{10}$. This example is also from \textcite{elvira2017dmmis} but with proper scaling so that the main support of $\pi$ is inside $[0,1]^{10}$. The mean $\mathbb{E}_{\pi}[X_j] = 0.550$ for $j = 1,\ldots,10$ and the normalizing constant $Z = 1$. We again use the Mean Squared Error (MSE) of the estimates for the performance evaluation. Similar to the experiment setup for the two dimensional problem in Subsection~\ref{subsec:simulation_pmc_2d}, we compare the PQMC to the PMC with and without the covariance adaptation. We run both PMC and PQMC for $T = 10$ iterations but vary $K$ and $J$ while keeping $KJ = 2{,}000$, so total of $TKJ = 20{,}000$ evaluations of the target distributions. The initial proposal centers are the $K$ Sobol' points over $[0,1]^{10}$. We use the same isotropic covariance matrix $\sigma^2 I_2$ for all the initial proposals and keep the adapted covariance isotropic. We compute the MSEs for both the standard PMC estimator and the weighted PMC estimator averaged over 100 independent runs. Table~\ref{tab:pmc_10d_full_m} shows the MSEs in log for the estimation of $\mathbb{E}_{\pi}[X]$. Similar to the conclusion drawn for the two-dimensional example, the weighted PMC estimator on PQMC samples outperforms all PMC settings for different values of $K$, $J$, and $\sigma$. Also, significant improvements are observed for PMC algorithms that have covariance adaptation, especially under the weighted PMC estimator. Table~\ref{tab:pmc_10d_full_z} in the Appendix shows the MSEs in log for the estimation of the normalizing constant $Z$. \par

\begin{table}[t!]
  \centering
  \resizebox{\columnwidth}{!}{%
  \begin{tabular}{|cccc|ccc|}
  \hline
Estimator & Algorithm & K & J & $\sigma = 0.1$ & $\sigma = 0.2$ & $\sigma = 0.5$ \\
\hline
Standard & PMC (Multinomial) & 50 & 40 & -4.95 [-8.68,-2.88] & -10.06 [-13.00,-7.35] & -6.66 [-8.34,-5.30] \\
Standard & PMC (Systematic) & 50 & 40 & -4.98 [-9.01,-2.74] & -10.04 [-11.98,-7.07] & -6.69 [-8.55,-5.48] \\
Standard & PMC (Multinomial + Lookback) & 50 & 40 & -8.66 [-11.93,-5.19] & -9.70 [-11.91,-6.08] & -8.71 [-11.12,-6.95] \\
Standard & PMC (Systematic + Lookback) & 50 & 40 & -7.86 [-11.94,-4.11] & -10.14 [-11.84,-8.31] & -8.66 [-11.38,-6.00] \\
Standard & PQMC (ISP + Lookback) & 50 & 40 & -8.90 [-12.31,-4.98] & -10.10 [-13.07,-7.40] & -8.85 [-11.26,-6.61] \\
\hdashline
Weighted & PMC (Multinomial) & 50 & 40 & -5.75 [-8.51,-3.36] & -11.00 [-12.44,-9.32] & -7.03 [-8.62,-5.74] \\
Weighted & PMC (Systematic) & 50 & 40 & -5.94 [-8.93,-3.28] & -11.17 [-12.95,-9.69] & -7.02 [-8.57,-5.96] \\
Weighted & PMC (Multinomial + Lookback) & 50 & 40 & -10.84 [-12.81,-9.31] & -10.91 [-12.80,-9.25] & -10.82 [-12.09,-9.37] \\
Weighted & PMC (Systematic + Lookback) & 50 & 40 & -10.77 [-12.69,-9.51] & -11.10 [-12.64,-9.67] & -10.85 [-12.29,-9.69] \\
Weighted & PQMC (ISP + Lookback) & 50 & 40 & \textbf{-12.06} [-13.92,-10.90] & \textbf{-12.13} [-13.63,-10.87] & \textbf{-11.95} [-13.91,-10.87] \\
\hline
Standard & PMC (Multinomial) & 100 & 20 & -5.32 [-8.05,-2.89] & -10.44 [-12.62,-8.19] & -6.67 [-8.01,-5.44] \\
Standard & PMC (Systematic) & 100 & 20 & -5.40 [-8.00,-2.73] & -10.74 [-12.84,-9.83] & -6.58 [-8.62,-4.73] \\
Standard & PMC (Multinomial + Lookback) & 100 & 20 & -6.96 [-12.08,-3.59] & -10.56 [-12.66,-8.82] & -8.43 [-10.61,-5.94] \\
Standard & PMC (Systematic + Lookback) & 100 & 20 & -7.69 [-12.36,-4.00] & -10.68 [-12.84,-8.50] & -8.26 [-10.87,-5.92] \\
Standard & PQMC (ISP + Lookback) & 100 & 20 & -8.97 [-12.16,-6.19] & -10.80 [-12.65,-7.81] & -8.62 [-11.76,-5.94] \\
\hdashline
Weighted & PMC (Multinomial) & 100 & 20 & -6.47 [-8.52,-4.55] & -11.31 [-12.99,-10.13] & -7.02 [-8.65,-5.64] \\
Weighted & PMC (Systematic) & 100 & 20 & -6.69 [-9.23,-4.40] & -11.33 [-12.79,-9.77] & -6.90 [-8.85,-5.33] \\
Weighted & PMC (Multinomial + Lookback) & 100 & 20 & -11.35 [-13.52,-9.49] & -11.41 [-13.58,-10.22] & -11.10 [-13.00,-9.81] \\
Weighted & PMC (Systematic + Lookback) & 100 & 20 & -11.42 [-12.85,-10.44] & -11.33 [-12.93,-10.34] & -11.21 [-12.89,-9.45] \\
Weighted & PQMC (ISP + Lookback) & 100 & 20 & \textcolor{red}{\textbf{-12.11}} [-13.55,-10.89] & \textcolor{red}{\textbf{-12.25}} [-13.58,-11.22] & \textcolor{red}{\textbf{-11.98}} [-13.49,-10.80] \\
\hline
  \end{tabular}
  }
  \caption{LogMSEs in the estimation of $\mathbb{E}_{\pi}[X]$ for the ten dimensional mixture of three normals using different values of $K$, $J$, and $\sigma$ with the initial proposal centers being the $K$ Sobol' points over $[0,1]^{10}$. The number of evaluations of the target distribution is fixed to $TKJ = 20{,}000$. The MSEs are averaged over 100 independent runs and shown in log under format ``mean [min,max]". The best results for each value of $\sigma$ are highlighted in red bold-face.}
  \label{tab:pmc_10d_full_m}
\end{table}

\begin{figure}[t!]
  \centering
  \includegraphics[width=0.9\textwidth]{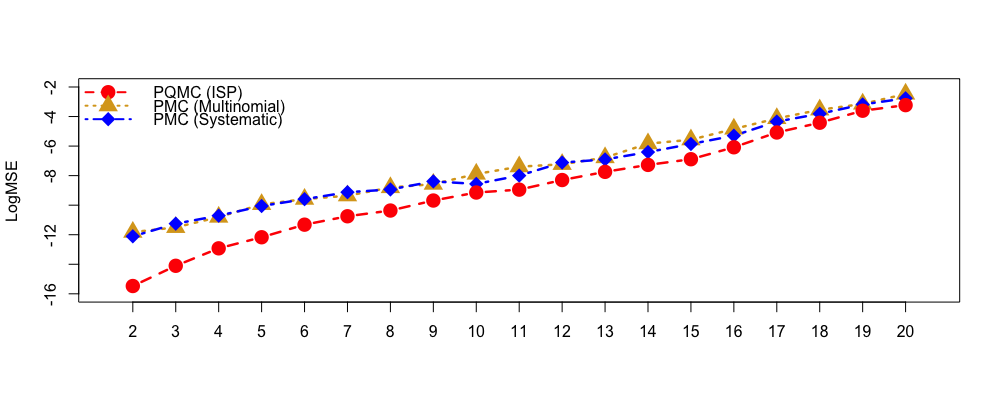}
  \caption{LogMSEs in the estimation of $Z$ for the mixture of three normals with $K = 50$, $J = 40$, and $T = 10$ for $p = 2,\ldots,20$. The initial proposal centers are the $K$ Sobol' points over $[0,1]^{p}$. The initial proposal covariances are $0.2^2 I_{p}$ and updated by lookback adaptation. The estimation is by the weighted PMC estimator. The MSEs are averaged over 100 independent runs.}
  \label{fig:pmc_hd2_z_comp}
\end{figure}

To better study the performance of PQMC as the dimension $p$ increases, we change the dimension for the mixture of three normals in \eqref{eq:mixture_10d} while keeping the same structure for the means and covariances. We compare the performance of PQMC to PMC with covariance adaptation. We run the algorithms for $T = 10$ iterations with $K = 50$ proposals and $J = 40$ samples simulated from each proposal. We use the same isotropic covariance matrix $0.2^2 I_{p}$ for the initial proposal covariances and keep the covariances isotropic after lookback adaptation. The estimation is by the weighted PMC estimator for its empirical improvement over the standard PMC estimator when there is adaptation for the proposal covariances. Figure~\ref{fig:pmc_hd2_z_comp} shows evolution of the MSEs in log for the estimation of the normalizing constant $Z = 1$ as the dimension increases. We can see that PQMC outperforms the PMC for all dimensions up to $p = 20$ in this example, but the improvement diminishes as dimension goes up. The diminishing improvement is more obvious for the estimation of the mean $\mathbb{E}_{\pi}[X]$ presented in Figure~\ref{fig:pmc_hd2_m_comp} in the Appendix.

\section{Expensive Posterior Example: Friction Drilling}
\label{sec:simulation_pmc_drilling}

\begin{figure}[t!]
  \centering
  
  \begin{subfigure}{0.3\textwidth}
    \centering
    \includegraphics[width=0.9\textwidth]{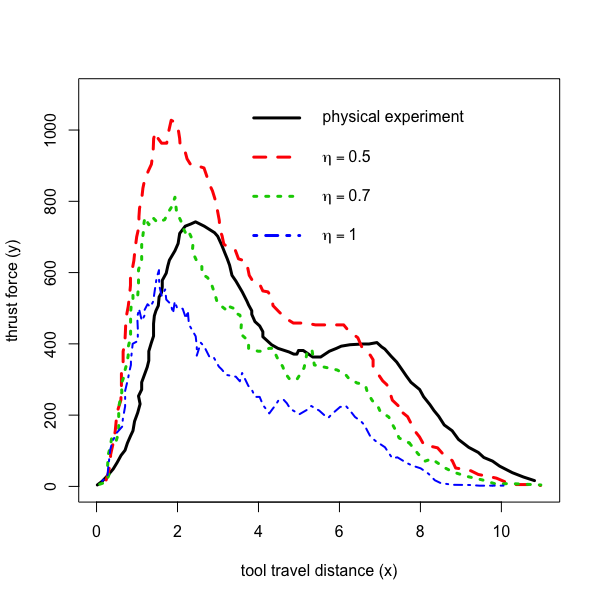}
    \caption{FEM}
  \end{subfigure}%
  \begin{subfigure}{0.3\textwidth}
    \centering
    \includegraphics[width=0.9\textwidth]{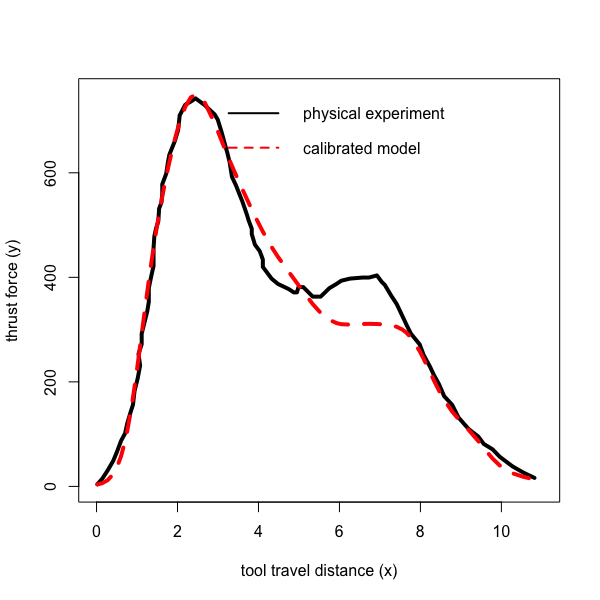}
    \caption{Calibrated}
  \end{subfigure}%
  
  \caption{Left Panel: FEM outputs at three values of the friction coefficient versus the physical experiment output. Right Panel: Calibrated model output at the posterior means, obtained by the weighted PMC estimators on PQMC samples, versus the physical experiment output.}
  \label{fig:drilling_output}
\end{figure}

\textcite{miller2007drilling} develop a thermomechanical finite element model (FEM) to simulate a fiction drilling process for analyzing the relationship between the thrust force ($y$) and the tool travel distance ($x$). There is an unknown parameter, the friction coefficient ($\eta$), in the FEM that one has to specify to obtain the FEM output. The left panel of Figure~\ref{fig:drilling_output} shows the FEM outputs of thrust force over the tool travel distance for three different values of the friction coefficient. A physical experiment is also performed to validate the FEM, where the actual experiment output is also presented in the left panel of Figure~\ref{fig:drilling_output}. From the plot, \textcite{miller2007drilling} conclude that $\eta = 0.7$ is the best choice for the coefficient of friction, but there is still a large discrepancy in the FEM predictions of the thrust force. A further investigation shows that due to the deflection of the sheet at the initial contact with the tool, the tool travel in the physical experiment is less than the tool travel inputted to the FEM, causing the discrepancy. However, fixing this in the FEM code is difficult and computationally expensive. \textcite{joseph2015calibration} propose an engineering-driven statistical adjustment that can reduce the discrepancy in a more efficient way. \par

Following the steps described in \textcite{joseph2019mined} Section 5, let $y = g(x;\eta)$ be the FEM, and introduce two adjustment parameters $\gamma_{1}$ and $\gamma_{2}$ such that $y = g(\gamma_{1}x^{\gamma_{2}};\eta)$ where $\gamma_{1}\in[0,1]$ accounts for the deflection at the initial contact and $\gamma_{2}$ reflects that the deflection could change during tool travel. $\gamma_{2} > 1$ indicates that a longer travel distance results in larger deflection. Thus, the calibration problem reduces to a nonlinear regression problem,
\begin{equation}
  \label{eq:drilling1}
  y_i = g(\gamma_{1}x_{i}^{\gamma_{2}};\eta) + \epsilon_{i}\; ,
\end{equation}
where $\epsilon_{i} \overset{\text{iid}}{\sim} \mathcal{N}(0,\sigma^2)$. However, the FEM $g(\cdot;\cdot)$ in the nonlinear regression is expensive to compute, so we approximate it using Gaussian Process (GP),
\begin{equation}
  \label{eq:drilling2}
  \hat{g}(x;\eta) = \exp\{\hat{\mu} + r(x;\eta)^{T}R^{-1}(\log y^{\text{FEM}} - \hat{\mu}1)\}\; ,
\end{equation}
where $\hat{\mu} = 1^{T}R^{-1}\log y^{\text{FEM}}/ 1^{T}R^{-1}1$. $r(x;\eta)$ is the correlation vector and $R$ is the correlation matrix both using the Gaussian correlation function $R(h) = \exp\{-\sum_{i}\theta_i h_i^2\}$. We use the R package \texttt{GPfit} \parencite{macdonald2015gpfitR} to fit the model. We use Bayesian inference to estimate the friction coefficient ($\eta$) and the two adjustment parameter ($\gamma_1,\gamma_2$), where the model is 
\begin{equation}
  \label{eq:drilling3}
  \begin{aligned}
    y_i &\overset{\text{iid}}{\sim} \mathcal{N}(\hat{g}(\gamma_{1}x_{i}^{\gamma_{2}};\eta), \sigma^2) \; \forall i = 1,\ldots,N \\
    \eta &\sim p(\eta; 0.5, 1, 10, 10) \\
    \gamma_1 &\sim p(\gamma_1; 0.5, 1, 10, 100) \\
    \gamma_2 &\sim p(\gamma_2; 0.75, 1.25, 10, 10) \\
    p(\sigma^{2}) &\propto 1 / \sigma^2
  \end{aligned}
\end{equation}
where $p(x;a,b,\lambda_{a},\lambda_{b}) = \exp\{\lambda_a(x-a)\}I(x<a) + I(a\leq x\leq b) + \exp\{-\lambda_b(x-b)\}I(x>b)$ is the prior distribution where Uniform prior is used for $x\in[a,b]$ and Exponential distribution is used for $x\notin[a,b]$. It follows that the posterior distribution is 
\begin{equation}
  \label{eq:drilling4}
  p(\eta,\gamma_{1},\gamma_{2},\sigma^2|y)\propto\frac{1}{\sigma^{N}}\exp\bigg\{-\frac{1}{2\sigma^2}\sum_{i=1}^{N}[y_i - \hat{g}(\gamma_{1}x_{i}^{\gamma_{2}};\eta)]^2\bigg\}\times p(\eta)p(\gamma_1)p(\gamma_2)p(\sigma^2) \; .
\end{equation}
We can integrate out $\sigma^2$, leading to the log posterior distribution,
\begin{equation}
  \label{eq:drilling5}
  \log p(\eta,\gamma_{1},\gamma_{2}|y) = \mbox{const.} - \frac{N}{2}\log\bigg(\sum_{i=1}^{N}[y_i - \hat{g}(\gamma_{1}x_{i}^{\gamma_{2}};\eta)]^2\bigg) + \log p(\eta) + \log p (\gamma_1) + \log p (\gamma_2)\; .
\end{equation}
There are 332 observations for the FEM, so one evaluation of the GP approximation is expensive, and we have to compute it for each of the $N = 96$ data points in the physical experiment. One evaluation of the posterior distribution takes more than 10 seconds on an average laptop. \par

We compare the performance of PQMC to PMC with covariance adaptation. We run both algorithms for $T = 7$ iterations with $K = 13$ proposals and $J = 7$ samples drawn from each proposal, thus leading to $TKJ =  637$ evaluations of the posterior distribution. The initial centers are the 13 Lattice points over $[0.5,1]\times[0.5,1]\times[0.75,1.25]$ that covers the key region of the prior, and the Minimax measure of the 13 points is 0.3 in the region $[0.5,1]\times[0.5,1]\times[0.75,1.25]$ computed using the \texttt{minimaxdesign} package in R \parencite{mak2019minimaxR}. Thus, we use the isotropic covariance matrix $0.2^2 I_{3}$ for all the initial proposals such that the proposals at the first iteration can cover up the main region of the prior, and we keep the adapted covariances isotropic. We compute the posterior means by the weighted PMC estimator. \par

\begin{figure}[t!]
  \centering
  
  \begin{subfigure}{0.3\textwidth}
    \centering
    \includegraphics[width=\textwidth]{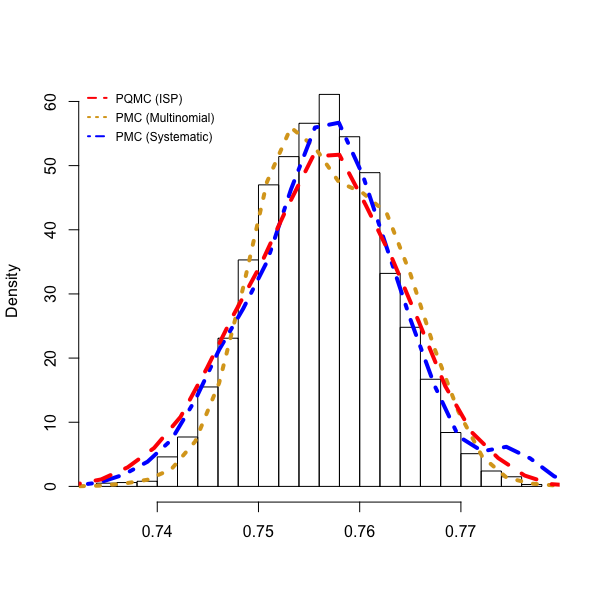}
    \caption{$\eta$}
  \end{subfigure}%
  \begin{subfigure}{0.3\textwidth}
    \centering
    \includegraphics[width=\textwidth]{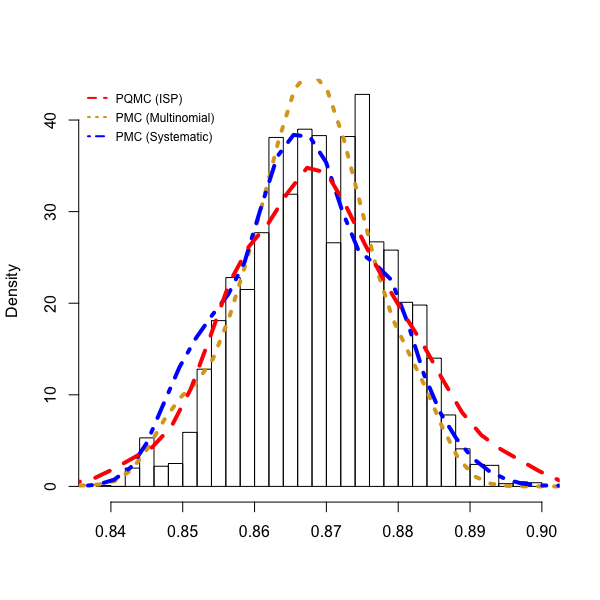}
    \caption{$\gamma_{1}$}
  \end{subfigure}%
  \begin{subfigure}{0.3\textwidth}
    \centering
    \includegraphics[width=\textwidth]{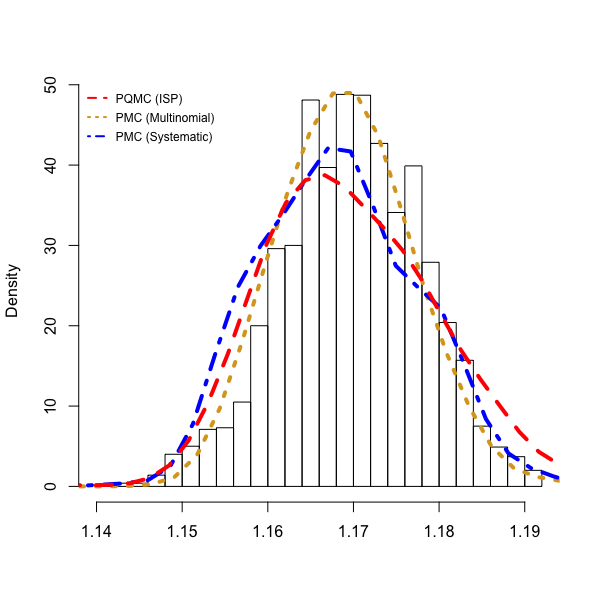}
    \caption{$\gamma_{2}$}
  \end{subfigure}%
  
  \caption{Histograms of the MCMC samples and the weighted marginal densities of the PQMC and PMC (multinomial and systematic) samples.}
  \label{fig:drilling_density}
\end{figure}

\begin{figure}[t!]
  \centering
  \includegraphics[width=0.75\textwidth]{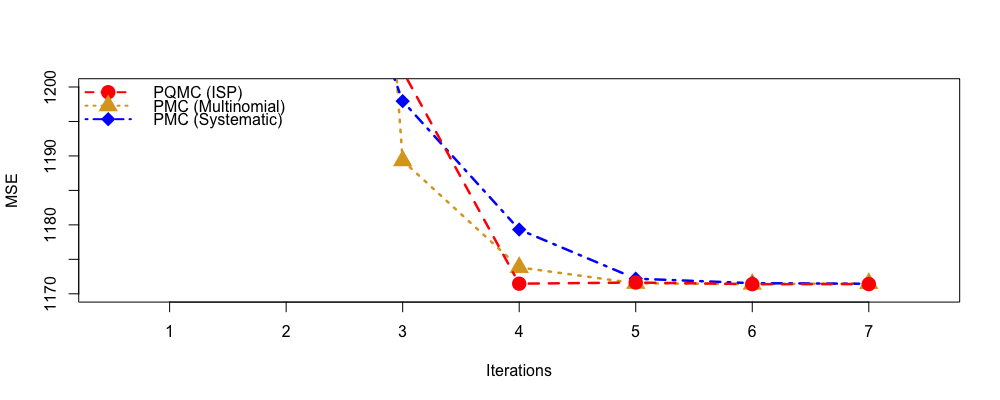}
  \caption{MSEs of the calibrated model predictions at posterior mean using PMC or PQMC samples up to $t$-th iteration to the physical experiment output.}
  \label{fig:drilling_convergence}
\end{figure}

The right panel of Figure~\ref{fig:drilling_output} shows the predictions from the calibrated model using the posterior means ($\hat{\eta} = 0.756$, $\hat{\gamma}_{1} = 0.869$, $\hat{\gamma}_{2} = 1.170$) computed from the PQMC samples. We can see that the calibration model helps bring the FEM model output much closer to the actual physical experiment data. The PMC with multinomial and systematic resampling both yield similar posterior means, and thus comparable calibrated models to the PQMC. Moreover, we also run the MCMC for 5000 iterations using normal proposal with covariance $0.01^2 I_3$. The starting point of the Markov Chain is at $[0.75,0.88,1.18]^{T}$, the posterior means computed in \textcite{joseph2019mined}. Figure~\ref{fig:drilling_density} shows the histograms of MCMC samples. The weighted marginal densities of the PQMC and PMC samples are plotted in Figure~\ref{fig:drilling_density}. The marginal density of the PQMC samples shows slightly better agreement with the MCMC samples overall. On the other hand, since the true posterior means cannot be computed analytically and it is also very expensive to approximate using the numerical integration method, we instead use the Mean Square Error, $\mbox{MSE} = N^{-1}(y_i - \hat{y}_i)^2$ where $\hat{y}_i = \hat{g}(\hat{\gamma}_{1}x_i^{\hat{\gamma}_2};\hat{\eta})$ is the prediction of the calibrated model at the posterior means, for evaluating the performance and convergence of the PMC and PQMC. Figure~\ref{fig:drilling_convergence} shows the MSEs of the calibrated model predictions at the posterior means constructed by weighted PMC estimator using samples up to the $t$-th iteration. Using the proposed MSE criterion, the PQMC converges in only 4 iterations ($4KJ = 364$ samples), while using PMC would require 5 iterations, demonstrating that faster convergence can be achieved by PQMC numerically.

\section{Conclusion}
\label{sec:conclusion}

This paper proposes the Population Quasi-Monte Carlo (Algorithm~\ref{algo:pqmc}) that incorporates Quasi-Monte Carlo ideas into the sampling and adaptation step of the generic Population Monte Carlo (Algorithm~\ref{algo:pmc}). For the \textit{sampling} step, we propose to use a set of random but low discrepancy points to replace the simple random samples from the proposal distributions. For the \textit{adaptation} step, we propose the importance support points (ISP) resampling, a deterministic resampling method that yields the set of resamples minimized over the energy distance to the original weighted samples such that most information can be retained. Numerical examples are shown to demonstrate the significant improvement of the ISP resampling over the traditional resampling methods for problems up to 20 dimensions. Given the Koksma-Hlawka-like bound presented by \textcite{mak2018sp} that connects the energy distance to the squared integration error from resampling, energy distance is a better measure for the effectiveness of resampling methods than the conditional variances shown in \eqref{eq:isprs2}. Within the PQMC framework, we also propose the lookback adaptation for updating the global covariance, where all proposals share the same covariance parameters. This adaptation is computationally efficient as it does not require additional evaluation of the proposal distribution. This covariance adaptation also demonstrates significant improvement when it is used in generic PMC with random resampling in numerical studies. This is especially important when the initial proposal covariances are chosen poorly, but this issue has received scant attention in the literature. Last, since there is adaptation in PMC and PQMC, the standard PMC estimator is not efficient, and we propose the weighted PMC estimator with the set of correction weights that are proportional to the effective samples size of each iteration. Extensive numerical studies in various settings presented in Section~\ref{sec:simulation} shows that PQMC yields faster convergence rate than the generic PMC, but more theoretical studies on PQMC are needed. \par

On the other hand, the ISP resampling in PQMC suffer more computational burden than the traditional resampling methods, as it requires $\mathcal{O}(M^2)$ evaluations to compute the pairwise distance for the weighted samples $\{(y_m,\bar{w}_m)\}_{m=1}^{M}$. However, in many real world Bayesian problems, the dominant computational cost is from the evaluation of the target distribution, as shown by the friction drilling calibration example in Section~\ref{sec:simulation_pmc_drilling}. Thus, it is justifiable to use a more computational expensive resampling scheme if it can result in faster convergence, hence reducing the evaluations of the target distribution while still achieving the desirable performance. On the other hand, as mentioned by \textcite{cornuet2012amis}, the initialization has a major impact on the performance for the class of adaptive importance sampling algorithm, as the adaptation is only based on the prior samples. It is difficult to recover from poor initialization as shown by the two dimensional mixture of five normals example with ``bad" initialized centers from $[0.4,0.6]^2$ and covariance being $0.1^2 I_2$. One promising solution is to allocate more resources at the initial stage of the PQMC by starting with a large $K$ (number of proposals) and slowly decreasing $K$ as the algorithm converges. Using ISP resampling can retain most information from the original samples and shows robust empirical performance even when $K$ is small, making it a perfect fit for the idea of using a decreasing sequence of $K$. It is an interesting future research direction to further explore. 

\bigskip

\printbibliography

\bigskip

\section*{\LARGE{Appendices}}

\appendix

\section{Importance Support Points}
\label{appendix:importance_support_points}

The importance support points aims to solve the following optimization problem,
\begin{equation}
  \label{eq:aisp1}
  \{\xi_{i}\}_{i=1}^{n} \in \arg\min_{x_1,\ldots,x_n\in\mathcal{X}}\frac{2}{n}\sum_{i=1}^{n}\sum_{m=1}^{M}\bar{w}_m\lVert x_i - y_m\rVert_{2} - \frac{1}{n^2}\sum_{i=1}^{n}\sum_{j=1}^{n}\lVert x_i - x_j\rVert_{2}\; ,
\end{equation}
where $\bar{w}_m$ is the normalized weight for the sample $y_m$. We can see that \eqref{eq:aisp1} is a difference of convex functions. One efficient algorithm for solving the difference of convex program in the literature is the convex-concave procedure (CCP) developed by \textcite{yuille2002ccp}. The main idea of CCP is to iterate over the convexification and optimization steps until convergence. The convexfication step is to replace the concave term in the objective with a convex upper bound. The optimization step is to solve the convex surrogate formulation using convex optimization methods. Following the same CCP for the Monte Carlo approximation \eqref{eq:sp2} of the support points in \textcite{mak2018sp}, we first majorize the concave term $n^{-2}\sum_{i=1}^{n}\sum_{j=1}^{n}\lVert x_i - x_j\rVert_{2}$ by the first-order Taylor expansion at the current iterate $\{x_{j}^{(t)}\}_{j=1}^{n}$, yielding the convex surrogate formulation, 
\begin{equation}
  \label{eq:aisp2}
  \begin{aligned}
  \arg\min_{x_1,\ldots,x_n\in\mathcal{X}} \quad &\frac{2}{n}\sum_{i=1}^{n}\sum_{m=1}^{M}w_m\lVert x_i - y_m\rVert_{2} - \\
  & \frac{1}{n^2}\bigg[\sum_{i=1}^{n}\sum_{j=1}^{n}\bigg(\lVert x_i^{(t)} - x_j^{(t)}\rVert_{2} + \frac{2(x_i - x_i^{(t)})^{T}(x_i^{(t)} - x_j^{(t)})}{\lVert x_i^{(t)} - x_j^{(t)}\rVert_{2}}\bigg)\bigg]\; .
  \end{aligned}
\end{equation}
In order to get a closed form solution for \eqref{eq:aisp2}, we further convexify it by the fact that $\frac{\lVert x\rVert_{2}^{2}}{2\lVert x^{(t)}\rVert_{2}}+\frac{\lVert x^{(t)}\rVert_{2}}{2}$ majorizes $\lVert x\rVert_{2}$ at $x^{(t)}$ for any $x^{(t)}\in\mathbb{R}^{p}$, leading to the following minimization,
\begin{equation}
  \label{eq:ispa3}
  \begin{aligned}
  \arg\min_{x_1,\ldots,x_n\in\mathcal{X}} \quad & \frac{2}{n}\sum_{i=1}^{n}\sum_{m=1}^{M}w_m\bigg[\frac{\lVert x_i - y_m\rVert_{2}^{2}}{2\lVert x_i^{(t)} - y_m\rVert_{2}} + \frac{\lVert x_i^{(t)} - y_m\rVert_{2}}{2} \bigg] - \\
  & \frac{1}{n^2}\bigg[\sum_{i=1}^{n}\sum_{j=1}^{n}\bigg(\lVert x_i^{(t)} - x_j^{(t)}\rVert_{2} + \frac{2(x_i - x_i^{(t)})^{T}(x_i^{(t)} - x_j^{(t)})}{\lVert x_i^{(t)} - x_j^{(t)}\rVert_{2}}\bigg)\bigg]\;,
  \end{aligned}
\end{equation}
where $\{x_i^{(t+1)}\}_{i=1}^{n}$ can now be obtained in closed form,
\begin{equation}
  \label{eq:ispa4}
  x_i^{(t+1)} = \frac{1}{\sum_{m=1}^{N}(w_m\lVert x_i^{(t)} - y_m\rVert_{2})^{-1}}\bigg[\frac{1}{n}\sum_{\substack{j=1\\j\neq i}}^{n}\frac{x_i^{(t)} - x_j^{(t)}}{\lVert x_i^{(t)} - x_j^{(t)}\rVert_{2}} + \sum_{m=1}^{M}w_m\frac{y_m}{\lVert x_i^{(t)} - y_m\rVert_{2}}\bigg]
\end{equation}
for $i = 1,\ldots,n$. Repeat the above until $\{x_i^{(t)}\}_{i=1}^{n}$ converges. Suppose the algorithm requires $T$ iterations till convergence, then the computational complexity is $\mathcal{O}(nMT)$. In practice, we generally require $T > 200$ for good performance. Moreover, focusing on the closed form update of \eqref{eq:ispa4}, we cannot have $x_i^{(t)} = y_m \; \forall t,i,m$; otherwise $\lVert x_i^{(t)} - y_m\rVert_{2} = 0$, and thus $\lVert x_i^{(t)} - y_m\rVert_{2}^{-1}$ is undefined. Thus, by using CCP, the resulted importance support points $\{\xi_{i}\}_{i=1}^{n}$ cannot be points from the original weighted samples $\{y_m\}_{m=1}^{M}$. That is why we need to resort to Algorithm~\ref{algo:isprs} for the ISP resampling. Moreover, the computational complexity of Algorithm~\ref{algo:isprs} is $\mathbb{O}(M^2)$. In the case of PMC, $K = n$ so $KJ = nJ = M$. $J$ is often not too large in PMC, so we can assume $J < T$, the number of iterations required for convergence in the above CCP approach. Thus, $M^2 = nMJ < nMT$, showing that sequential optimization procedure is actually more computational efficient than the CCP approach in the context of PMC with moderate $J$.

\section{Unbiasedness and Consistency of PMC Estimator}
\label{appendix:estimator}

\subsection{Importance Sampling Estimator}
\label{appendix:is}
Let $\pi = \gamma / Z$ be a probability density function on $\emptyset\neq\mathcal{X}\subseteq\mathbb{R}^{p}$ where $\gamma:\mathcal{X}\to\mathbb{R}^{+}$ is an nonnegative function that is known pointwise and $Z = \int_{\mathcal{X}}\gamma(x) dx$ is the unknown finite normalizing constant. In Bayesian inference problems, we are interested in solving 
\begin{equation}
  \label{eq:is1}
  I = \mathbb{E}_{\pi}[h(X)] = \int_{\mathcal{X}}h(x)\pi(x)dx = \int_{\mathcal{X}}h(x)\frac{\gamma(x)}{Z} dx = \frac{\int_{\mathcal{X}}h(x)\gamma(x)dx}{Z} = \frac{\int_{\mathcal{X}}h(x)\gamma(x)dx}{\int_{\mathcal{X}}\gamma(x)dx}
\end{equation}
for any function $h:\mathcal{X}\to\mathbb{R}$ that is integrable with respect to $\pi$. The key of Importance Sampling is to sample from another distribution $q$, the importance distribution, with support $\mathcal{X}$, then the above integral can be rewritten as an expectation with respect to $q$, 
\begin{equation}
  \label{eq:is2}
  I = \frac{\int_{\mathcal{X}}h(x)\gamma(x)dx}{\int_{\mathcal{X}}\gamma(x)dx} = \frac{\int_{\mathcal{X}}h(x)\frac{\gamma(x)}{q(x)}q(x)dx}{\int_{\mathcal{X}}\frac{\gamma(x)}{q(x)}q(x)dx} = \frac{\int_{\mathcal{X}}h(x)w(x)q(x)dx}{\int_{\mathcal{X}}w(x)q(x)dx} = \frac{\mathbb{E}_{q}[h(X)w(X)]}{\mathbb{E}_{q}[w(X)]}\; ,
\end{equation}
where $w(\cdot) = \gamma(\cdot) / q(\cdot)$ is the unnormalized importance weight function. By drawing $N$ particles $\{X_n\}_{n=1}^{N}$ from the importance distribution $q$, the IS estimator is 
\begin{equation}
  \label{eq:is3}
  \hat{I}_N = \frac{\frac{1}{N}\sum_{n=1}^{N}h(X_n)w(X_n)}{\frac{1}{N}\sum_{n=1}^{N}w(X_n)} \overset{p}{\to} \frac{\mathbb{E}_{q}[h(X)w(X)]}{\mathbb{E}_{q}[w(X)]} = I\; ,
\end{equation}
which is consistent by the Law of Large Numbers and Slutzky's theorem. Applying Delta method, 
\begin{equation}
  \label{eq:is4}
  \sqrt{N}(\hat{I}_{N} - I) \overset{d}{\to} N\left(0, \frac{\mathbb{E}_{q}[w^2(X)(h(X) - I)^2]}{\mathbb{E}_{q}[w(X)]^2}\right)\; .
\end{equation} 
Focusing on the numerator of the variance, 
\begin{equation}
  \label{eq:is5}
  \mathbb{E}_{q}[w^2(X)(h(X) - I)^2] = \int_{\mathcal{X}}\frac{\gamma^2(x)}{q^2(x)}(h(x) - I)^2 q(x) dx = \int_{\mathcal{X}}\frac{\gamma^2(x)(h(x) - I)^2}{q(x)} dx\; .
\end{equation} 
If $q$ has thinner tail than $\gamma$, the above integral could be infinite, so $q$ needs to have a fatter tail than $\gamma$. Also, the variance is large if $q$ is small over region where $\gamma$ is large, thus we also want $q$ and $\gamma$ to be similar in shape. Moreover, the variance also depends on the integrand $h$. Effective sample size is another diagnostic metric for evaluating IS that is free of the integrand $h$. Consider the general form of a weighted estimator, $\hat{I} = (\sum_{i=1}^{N}w_i Y_i)/(\sum_{i=1}^{N}w_i)$ where $Y_i$'s are independent random variables with same mean and variance $\sigma^2 > 0$ and $w_i\in[0,\infty)$ is the weight for $Y_i$. If $Y_i$ are unweighted, i.e. $w_1 = \cdots = w_N$, then the variance of $\hat{I}$ is $\sigma^2/N$. The effective sample size $N_e$ is the number of unweighted samples that achieves the variance of the weighted estimator,
\begin{equation}
  \label{eq:is6}
  \frac{\sigma^2}{N_e} = \mathbb{V}[\hat{I}] = \frac{\sum_{i=1}^{N}w_i^2}{(\sum_{i=1}^{N}w_i)^2}\sigma^2 \quad \Rightarrow \quad N_e = \frac{(\sum_{i=1}^{N}w_i)^2}{\sum_{i=1}^{N}w_i^2}\; .
\end{equation}
Thus, if the weights are too imbalanced, the effective sample size $N_e$ will be much smaller than $N$, resulting in an unreliable estimator. Given the limitation that finding a good importance distribution is difficult especially when the target distribution is complex high dimensional, standard IS is barely used in practice, but it is the key ingredient for PMC and SMC.

\subsection{Multiple Importance Sampling Estimator}
\label{appendix:mis}

Instead of entrusting the performance to a single proposal distribution, one improvement is to use a population of proposal distributions $\{q_k\}_{k=1}^{K}$ \parencite{elvira2019mis}. This is known as the Multiple Importance Sampling (MIS). Let us first assume that the normalizing constant $Z$ is known. Consider that $J$ samples are drawn from each proposal, i.e. $x_{k,j} \sim q_k$, we show that both the standard weighting and deterministic mixture weighting yield the unbiased estimator for $I = \mathbb{E}_{\pi}[h(X)]$. For the standard weighting scheme, $w_{k,j} = \gamma(x_{k,j}) / q_k(x_{k,j})$, the MIS estimator is
\begin{equation}
  \label{eq:mis_st_u1}
  \hat{I}^{\text{ST}} = \frac{1}{Z}\bigg(\frac{1}{KJ}\sum_{k=1}^{K}\sum_{j=1}^{J}h(x_{k,j})w_{k,j}\bigg) = \frac{1}{Z}\bigg(\frac{1}{KJ}\sum_{k=1}^{K}\sum_{j=1}^{J}h(x_{k,j})\frac{\gamma(x_{k,j})}{q_k(x_{k,j})}\bigg)\; .
\end{equation}
Since $x_{k,j}\sim q_k$, it follows that 
\begin{equation}
  \label{eq:mis_st_u2}
  \mathbb{E}[\hat{I}^{\text{ST}}] = \frac{1}{KJ}\sum_{k=1}^{K}\sum_{j=1}^{J}\int_{\mathcal{X}}h(x)\frac{\gamma(x)/Z}{q_k(x)}q_k(x)dx = \frac{1}{KJ}\sum_{k=1}^{K}\sum_{j=1}^{J}\int_{\mathcal{X}}h(x)\pi(x)dx = I\; .
\end{equation}
For the deterministic mixture weight, $w_{k,j} = \gamma(x_{k,j})/[K^{-1}\sum_{i=1}^{K}q_i(x_{k,j})]$, the MIS estimator is 
\begin{equation}
  \label{eq:mis_dm_u1}
  \hat{I}^{\text{DM}} = \frac{1}{Z}\bigg(\frac{1}{KJ}\sum_{k=1}^{K}\sum_{j=1}^{J}h(x_{k,j})w_{k,j}\bigg) = \frac{1}{Z}\bigg(\frac{1}{KJ}\sum_{k=1}^{K}\sum_{j=1}^{J}h(x_{k,j})\frac{\gamma(x_{k,j})}{K^{-1}\sum_{i=1}^{K}q_i(x_{k,j})}\bigg)\; .
\end{equation}
Since $x_{k,j}\sim q_k$, it follows that 
\begin{equation}
  \label{eq:mis_dm_u2}
  \begin{aligned}
  \mathbb{E}[\hat{I}^{\text{DM}}] &= \frac{1}{KJ}\sum_{k=1}^{K}\sum_{j=1}^{J}\int_{\mathcal{X}}h(x)\frac{\gamma(x)/Z}{K^{-1}\sum_{i=1}^{K}q_i(x)}q_k(x)dx \\
  &= \frac{1}{J}\sum_{j=1}^{J}\int_{\mathcal{X}}h(x)\frac{\pi(x)}{K^{-1}\sum_{i=1}^{K}q_i(x)}\bigg[K^{-1}\sum_{k=1}^{K}q_k(x)\bigg]dx \\
  &= I\; .
  \end{aligned}
\end{equation}
Now suppose that the normalizing constant $Z$ is unknown. We use a consistent estimator to replace $Z$ in both $\hat{I}^{\text{ST}}$ and $\hat{I}^{\text{DM}}$. Moreover, let us assume that $q_k$'s are all independent and have heavier tails than $\pi$ so the variance of the estimator is finite. For the standard weighting scheme, $w_{k,j} = \gamma(x_{k,j}) / q_k(x_{k,j})$, the consistent estimator for $Z$ is 
\begin{equation}
  \label{eq:mis_st_c1}
  \hat{Z}^{\text{ST}} = \frac{1}{KJ}\sum_{k=1}^{K}\sum_{j=1}^{J}w_{k,j} = \frac{1}{K}\sum_{k=1}^{K}\bigg[\frac{1}{J}\sum_{j=1}^{J}\frac{\gamma(x_{k,j})}{q_k(x_{k,j})}\bigg] = \frac{1}{K}\sum_{k=1}^{K}\hat{Z}^{\text{ST}}_k\; ,
\end{equation}
where $\hat{Z}^{\text{ST}}_k = J^{-1}\sum_{j=1}^{J}\gamma(x_{k,j})/q_k(x_{k,j})$. Consider the case when $J\to\infty$, $\hat{Z}^{\text{ST}}_k\overset{p}{\to}Z \; \forall k$ by standard IS argument, Thus, $\hat{Z}^{\text{ST}} = K^{-1}\sum_{k=1}^{K}\hat{Z}^{\text{ST}}_k \overset{p}{\to} Z$ by the Slutzky's theorem. Next consider when $K\to\infty$, one can verify that $\mathbb{E}[\hat{Z}^{\text{ST}}_k] = Z$, and thus $\mathbb{E}[\hat{Z}^{\text{ST}}] = Z$. It follows that for any $\epsilon > 0$,
\begin{equation}
  \label{eq:mis_st_c2}
  \lim_{K\to\infty}P(|\hat{Z}^{\text{ST}} - Z| > \epsilon) \leq \lim_{K\to\infty}\frac{\mathbb{V}[K^{-1}\sum_{k=1}^{K}\hat{Z}^{\text{ST}}_k]}{\epsilon^2} = \lim_{K\to\infty}\frac{\sum_{k=1}^{K}\mathbb{V}[\hat{Z}^{\text{ST}}_k]}{\epsilon^2K^2} = 0
\end{equation}
by the Chebyshev's inequality. Thus, $\hat{Z}^{\text{ST}}$ is a consistent estimator for $Z$ when either $K$ or $J$ goes to infinity. Similar argument can be provided to show that 
\begin{equation}
  \label{eq:mis_st_c3}
  \frac{1}{KJ}\sum_{k=1}^{K}\sum_{j=1}^{J}h(x_{k,j})\frac{\gamma(x_{k,j})}{q_k(x_{k,j})}\overset{p}{\to} ZI
\end{equation}
and by the Slutzky's theorem, it follows that
\begin{equation}
  \label{eq:mis_st_c4}
  \hat{I}^{\text{ST}} = \frac{1}{\hat{Z}^{\text{ST}}}\bigg(\frac{1}{KJ}\sum_{k=1}^{K}\sum_{j=1}^{J}h(x_{k,j})\frac{\gamma(x_{k,j})}{q_k(x_{k,j})}\bigg) \overset{p}{\to} I
\end{equation}
is consistent when the normalizing constant $Z$ is unknown. \par

Next consider the deterministic mixture weighting scheme $w_{k,j} = \gamma(x_{k,j})/[K^{-1}\sum_{i=1}^{K}q_i(x_{k,j})]$. Since exactly $J$ samples are drawn from each proposal, the $K$ proposals can be considered as one proposal that is $K$-component mixture with equal weights. Thus, following the standard IS argument, as $KJ\to\infty$,
\begin{equation}
  \label{eq:mis_dm_c1}
  \frac{1}{\hat{Z}^{\text{DM}}}\bigg(\frac{1}{KJ}\sum_{k=1}^{K}\sum_{j=1}^{J}h(x_{k,j})\frac{\gamma(x_{k,j})}{K^{-1}\sum_{i=1}^{K}q_i(x_{k,j})}\bigg) \overset{p}{\to} I\; ,
\end{equation}
where 
\begin{equation}
  \label{eq:mis_dm_c2}
  \hat{Z}^{\text{DM}} = \frac{1}{KJ}\sum_{k=1}^{K}\sum_{j=1}^{J}\frac{\gamma(x_{k,j})}{K^{-1}\sum_{i=1}^{K}q_i(x_{k,j})} \overset{p}{\to} Z\; .
\end{equation}

\subsection{Standard PMC Estimator}
\label{appendix:pmc}

A further refinement can be achieved by performing adaptation on the proposals: we iteratively draw samples from the proposals and use the samples to adapt the proposals' parameters to reduce the mismatch between the proposal and target distribution \parencite{bugallo2017ais}. It is known as the Adaptive Importance Sampling (AIS). Population Monte Carlo (PMC) by \textcite{cappe2004pmc} offers a flexible framework for combining AIS and MIS. Again, let us first assume that $Z$ is known. Suppose there are $K$ proposals $\{q_k^{(t)}\}_{k=1}^{K}$ at each iteration and $J$ samples are simulated from each proposal, i.e. $x_{k,j}^{(t)} \sim q_k^{(t)}$, with the deterministic mixture weighting scheme, the $T$ steps standard PMC estimator is 
\begin{equation}
  \label{eq:pmc_u1}
  \hat{I}^{\text{PMC}} = \frac{1}{Z}\bigg(\frac{1}{TKJ}\sum_{t=1}^{T}\sum_{k=1}^{K}\sum_{j=1}^{J}h(x_{k,j}^{(t)})\frac{\gamma(x_{k,j}^{(t)})}{K^{-1}\sum_{i=1}^{K}q_{i}^{(t)}(x_{k,j}^{(t)})}\bigg) = \frac{1}{T}\sum_{t=1}^{T}\hat{I}^{\text{PMC}}_{t}
\end{equation}
where 
\begin{equation}
  \label{eq:pmc_u2}
  \hat{I}^{\text{PMC}}_{t} = \frac{1}{Z}\bigg(\frac{1}{KJ}\sum_{k=1}^{K}\sum_{j=1}^{J}h(x_{k,j}^{(t)})\frac{\gamma(x_{k,j}^{(t)})}{K^{-1}\sum_{i=1}^{K}q_{i}^{(t)}(x_{k,j}^{(t)})}\bigg)
\end{equation}
is the MIS estimator with deterministic mixture weighting scheme on the $t$-th iteration's samples. Thus, the standard PMC estimator is the average of $T$ MIS estimators. Using the same argument for showing unbiasedness of the MIS estimator, $\mathbb{E}[\hat{I}^{\text{PMC}}_{t}] = I\;\forall t$, and thus $\mathbb{E}[\hat{I}^{\text{PMC}}] = I$. When the normalizing constant $Z$ is unknown, we replace $Z$ in $\hat{I}^{\text{PMC}}$ by a consistent estimator,
\begin{equation}
  \label{eq:pmc_c1}
  \hat{Z}^{\text{PMC}} = \frac{1}{TKJ}\sum_{t=1}^{T}\sum_{k=1}^{K}\sum_{j=1}^{J}\frac{\gamma(x_{k,j}^{(t)})}{K^{-1}\sum_{i=1}^{K}q_{i}^{(t)}(x_{k,j}^{(t)})} = \frac{1}{T}\sum_{t=1}^{T}\hat{Z}^{\text{PMC}}_{t}\; ,
\end{equation}
where 
\begin{equation}
  \label{eq:pmc_c2}
  \hat{Z}^{\text{PMC}}_{t} = \frac{1}{KJ}\sum_{k=1}^{K}\sum_{j=1}^{J}\frac{\gamma(x_{k,j}^{(t)})}{K^{-1}\sum_{i=1}^{K}q_{i}^{(t)}(x_{k,j}^{(t)})}
\end{equation}
is the MIS estimator of the normalizing constant $Z$ using the samples simulated at the $t$-th iteration of PMC. Following the same argument for proving the consistent of the MIS estimator, $\hat{Z}^{\text{PMC}}_{t}\overset{p}{\to} Z$ as $KJ \to \infty$, and thus $\hat{Z}^{\text{PMC}}\overset{p}{\to} Z$ by the Slutzky's theorem. It follows that 
\begin{equation}
  \label{eq:pmc_c3}
  \hat{I}^{\text{PMC}} = \frac{1}{\hat{Z}^{\text{PMC}}}\bigg(\frac{1}{TKJ}\sum_{t=1}^{T}\sum_{k=1}^{K}\sum_{j=1}^{J}h(x_{k,j}^{(t)})\frac{\gamma(x_{k,j}^{(t)})}{K^{-1}\sum_{i=1}^{K}q_{i}^{(t)}(x_{k,j}^{(t)})}\bigg) \overset{p}{\to} I\; .
\end{equation}

\subsection{Weighted PMC Estimator}
\label{appendix:wpmc}

Since there is adaptation in the PMC algorithm, the standard PMC estimator is not efficient as it treats samples across different iterations equally, but the samples at the early stages might be bad comparatively to the samples simulated from the adapted proposals. Weighted PMC estimator provides a way to ``forget" the poor samples by introducing a set of correction weights $\{\alpha^{(t)}\}_{t=1}^{T}$ with the constraint that $\sum_{t=1}^{T}\alpha^{(t)} = 1$. The weighted PMC (WPMC) estimator is
\begin{equation}
  \label{eq:wpmc_a1}
  \hat{I}^{\text{WPMC}} = \sum_{t=1}^{T}\alpha^{(t)}\hat{I}^{\text{PMC}}_t = \frac{1}{Z}\bigg(\frac{1}{KJ}\sum_{t=1}^{T}\sum_{k=1}^{K}\sum_{j=1}^{J}\alpha^{(t)}h(x_{k,j}^{(t)})\frac{\gamma(x_{k,j}^{(t)})}{K^{-1}\sum_{i=1}^{K}q_{i}^{(t)}(x_{k,j}^{(t)})}\bigg)
\end{equation}
when the normalizing constant $Z$ is known. If $Z$ is unknown, we replace it by a consistent estimator 
\begin{equation}
  \label{eq:wpmc_a2}
  \hat{Z}^{\text{WPMC}} = \sum_{t=1}^{T}\alpha^{(t)}\hat{Z}^{\text{PMC}}_t = \frac{1}{KJ}\sum_{t=1}^{T}\sum_{k=1}^{K}\sum_{j=1}^{J}\alpha^{(t)}\frac{\gamma(x_{k,j}^{(t)})}{K^{-1}\sum_{i=1}^{K}q_{i}^{(t)}(x_{k,j}^{(t)})}\; .
\end{equation}
Same argument for showing the unbiasedness and consistency of the standard PMC estimator can also be applied here to prove the unbiasedness (when $Z$ is known) and consistency (when $Z$ is unknown) for the weighted PMC estimator. Let us further approximate the weighted PMC estimator as the following,
\begin{equation}
  \label{eq:wpmc_a3}
  \hat{I}^{\text{WPMC}} = \sum_{t=1}^{T}\alpha^{(t)}\hat{I}^{\text{PMC}}_t \approx \sum_{t=1}^{T}\alpha^{(t)}\tilde{I}^{\text{PMC}}_t\; ,
\end{equation}
where 
\begin{equation}
  \label{eq:wpmc_a4}
  \tilde{I}^{\text{PMC}}_t = \frac{1}{\hat{Z}^{\text{PMC}}_{t}}\bigg(\frac{1}{KJ}\sum_{k=1}^{K}\sum_{j=1}^{J}h(x_{k,j}^{(t)})\frac{\gamma(x_{k,j}^{(t)})}{K^{-1}\sum_{i=1}^{K}q_{i}^{(t)}(x_{k,j}^{(t)})}\bigg)
\end{equation}
where the normalizing constant $Z$ is replaced by the MIS estimator $\hat{Z}^{\text{PMC}}_{t}$ constructing from the $t$-th iteration samples only instead of by the $\hat{Z}^{\text{WPMC}}$ from all PMC samples. $\tilde{I}^{\text{PMC}}_t$ is also the self-normalized MIS estimator on the $t$-th iteration samples. Assuming that $\{\tilde{I}^{\text{PMC}}_{t}\}_{t=1}^{T}$ are independent, the variance of the weighted PMC estimator can be approximated by
\begin{equation}  
  \label{eq:wpmc_a5}
  \mathbb{V}[\hat{I}^{\text{WPMC}}] \approx \sum_{t=1}^{T}(\alpha^{(t)})^2\mathbb{V}[\tilde{I}^{\text{PMC}}_{t}]\; .
\end{equation}
Thus, we want to find the correction weights $\alpha^{(t)}$ such that the variance of the weighted PMC estimator can be minimized. From \textcite{douc2007dmis2}, the optimal $\alpha^{(t)} \propto (\mathbb{V}[\tilde{I}^{\text{PMC}}_{t}])^{-1}$. However, the variance of $\tilde{I}^{\text{PMC}}_{t}$ depends on the integrand $h$, so we want to find an approximation that is free of the integrand. Let $N_{e}^{(t)}$ denotes the effective sample size of the $t$-th iteration samples, then $\mathbb{V}[\tilde{I}^{\text{PMC}}_{t}] \approx \sigma^2_{h}/N_{e}^{(t)}$ where $\sigma^2_h$ is the variance from the integrand $h$. Thus, the optimal correction weights can be computed by
\begin{equation}  
  \label{eq:wpmc_a6}
  \alpha^{(t)} = N_{e}^{(t)} / \sum_{i=1}^{T} N_{e}^{(i)}
\end{equation}
that is free from the integrand $h$. This approach does not require knowing the normalizing constant of the target distribution, but when all the unnormalized weights are very small, then it could lead to an undesirable large effective sample size $N_{e}^{(t)}$ as pointed out in \textcite{portier2018ais}. However, this rarely occurs in practice if the initial proposals are chosen appropriately. 

\section{Additional Simulation Results}
\label{appendix:simulation}

\begin{figure}[H]
  \centering
  \includegraphics[width=0.9\textwidth]{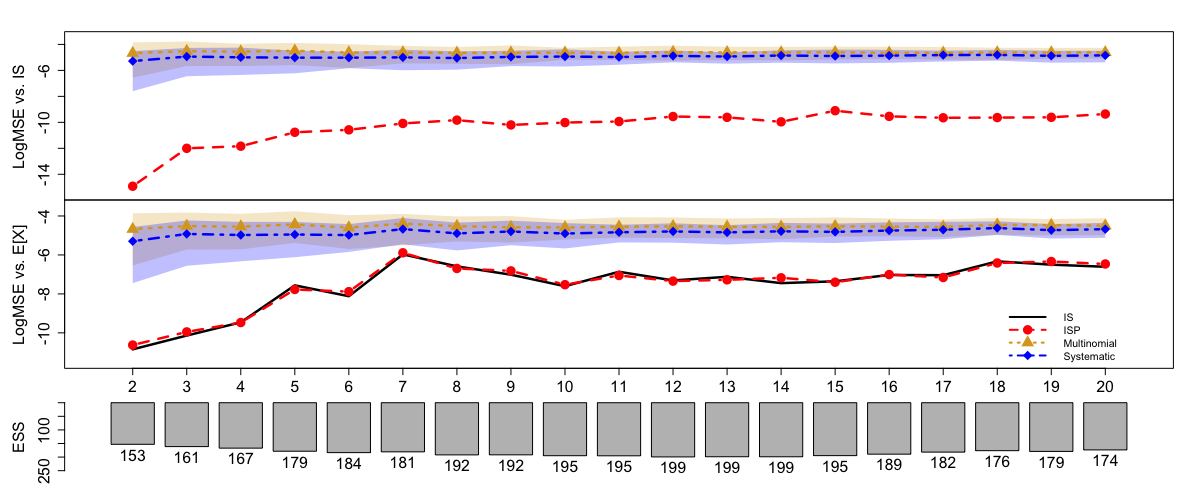}
  \caption{LogMSEs in the estimation of the IS estimator and $\mathbb{E}[X]$ where $X\sim\mathcal{N}(0,1)^{p}$ for $p = 2,\ldots,20$ using 100 resampled points from the 1{,}000 inverse Sobol' points of $q = \mathcal{N}(0,3^{(2/p^{0.8})})^{p}$ as the importance samples by multinomial, systematic, and ISP resampling. MSE for multinomial and systematic are averaged over 100 independent runs. ESS stands for effective sample size. Lines denote the logMSEs, and shaded bands mark the 10th and 90th quantiles.}
  \label{fig:resample_rhd}
\end{figure}

\begin{figure}[H]
  \centering
  \includegraphics[width=0.9\textwidth]{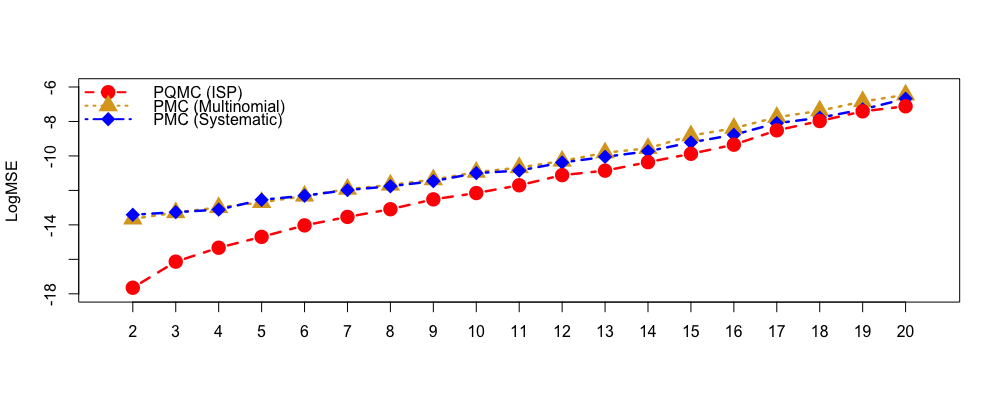}
  \caption{LogMSEs in the estimation of $\mathbb{E}_{\pi}[X]$ for the mixture of three normals with $K = 50$, $J = 40$, and $T = 10$ for $p = 2,\ldots,20$. The initial proposal centers are the $K$ Sobol' points over $[0,1]^{p}$. The initial proposal covariances are $0.2^2 I_{p}$ and updated by lookback adaptation. The estimation is by the weighted PMC estimator. The MSEs are averaged over 100 independent runs.}
  \label{fig:pmc_hd2_m_comp}
\end{figure}

\begin{table}[H]
  \centering
  \resizebox{\columnwidth}{!}{%
  \begin{tabular}{|cccc|ccc|}
  \hline
Estimator & Algorithm & K & J & $\sigma = 0.1$ & $\sigma = 0.2$ & $\sigma = 0.5$ \\
\hline
Standard & PMC (Multinomial) & 25 & 40 & -5.82 [-20.38,-3.36] & -5.98 [-20.33,-4.04] & -5.30 [-17.65,-3.53] \\
Standard & PMC (Systematic) & 25 & 40 & -6.65 [-15.85,-3.46] & -6.29 [-15.68,-4.19] & -5.25 [-15.10,-3.43] \\
Standard & PMC (Multinomial + Lookback) & 25 & 40 & -5.36 [-16.46,-3.15] & -5.56 [-21.06,-3.03] & -6.17 [-15.92,-4.11] \\
Standard & PMC (Systematic + Lookback) & 25 & 40 & -6.24 [-20.01,-3.64] & -6.16 [-17.12,-3.74] & -6.20 [-19.16,-3.44] \\
Standard & PQMC (ISP + Lookback) & 25 & 40 & -9.81 [-20.09,-7.93] & -8.67 [-15.93,-6.83] & -7.18 [-17.89,-5.37] \\
\hdashline
Weighted & PMC (Multinomial) & 25 & 40 & -5.77 [-18.29,-3.25] & -5.89 [-16.13,-4.05] & -5.27 [-14.14,-3.42] \\
Weighted & PMC (Systematic) & 25 & 40 & -6.63 [-17.71,-3.56] & -6.32 [-15.34,-4.30] & -5.08 [-15.44,-3.14] \\
Weighted & PMC (Multinomial + Lookback) & 25 & 40 & -5.18 [-16.28,-3.18] & -5.32 [-15.15,-2.63] & -5.70 [-17.70,-3.36] \\
Weighted & PMC (Systematic + Lookback) & 25 & 40 & -5.92 [-30.82,-3.12] & -5.82 [-20.23,-3.15] & -5.95 [-18.70,-3.36] \\
Weighted & PQMC (ISP + Lookback) & 25 & 40 & \textcolor{red}{\textbf{-12.42}} [-24.22,-10.60] & \textcolor{red}{\textbf{-12.09}} [-17.51,-10.26] & \textcolor{red}{\textbf{-11.37}} [-18.90,-9.50] \\
\hline
Standard & PMC (Multinomial) & 50 & 20 & -7.31 [-19.35,-5.02] & -6.23 [-14.11,-4.48] & -5.45 [-16.01,-3.49] \\
Standard & PMC (Systematic) & 50 & 20 & -7.43 [-19.36,-5.42] & -6.52 [-18.26,-4.85] & -5.52 [-12.28,-3.26] \\
Standard & PMC (Multinomial + Lookback) & 50 & 20 & -8.33 [-16.21,-6.31] & -7.49 [-19.55,-5.65] & -6.52 [-18.94,-4.32] \\
Standard & PMC (Systematic + Lookback) & 50 & 20 & -8.45 [-17.11,-6.57] & -7.62 [-18.01,-5.48] & -7.00 [-15.55,-5.13] \\
Standard & PQMC (ISP + Lookback) & 50 & 20 & -9.31 [-16.82,-7.02] & -8.30 [-21.69,-6.18] & -6.92 [-16.52,-5.08] \\
\hdashline
Weighted & PMC (Multinomial) & 50 & 20 & -7.41 [-16.52,-5.06] & -6.16 [-16.64,-4.21] & -5.34 [-13.08,-3.01] \\
Weighted & PMC (Systematic) & 50 & 20 & -7.47 [-16.60,-5.67] & -6.43 [-12.95,-4.68] & -5.38 [-12.21,-2.70] \\
Weighted & PMC (Multinomial + Lookback) & 50 & 20 & -9.21 [-18.38,-7.02] & -9.11 [-20.17,-6.79] & -7.63 [-19.57,-3.41] \\
Weighted & PMC (Systematic + Lookback) & 50 & 20 & -9.58 [-20.39,-7.67] & -9.50 [-17.17,-7.58] & -8.72 [-18.37,-6.60] \\
Weighted & PQMC (ISP + Lookback) & 50 & 20 & \textbf{-12.19} [-22.25,-9.91] & \textbf{-11.66} [-18.95,-9.66] & \textbf{-10.66} [-18.83,-8.08] \\
\hline
Standard & PMC (Multinomial) & 100 & 10 & -7.40 [-17.66,-5.01] & -6.50 [-16.71,-4.46] & -5.22 [-14.03,-3.35] \\
Standard & PMC (Systematic) & 100 & 10 & -7.60 [-15.38,-5.59] & -6.67 [-16.19,-4.68] & -5.54 [-16.56,-3.24] \\
Standard & PMC (Multinomial + Lookback) & 100 & 10 & -8.34 [-22.13,-6.24] & -7.74 [-19.12,-5.68] & -6.71 [-16.08,-4.31] \\
Standard & PMC (Systematic + Lookback) & 100 & 10 & -8.45 [-17.44,-6.01] & -7.89 [-18.95,-5.60] & -7.01 [-14.37,-5.16] \\
Standard & PQMC (ISP + Lookback) & 100 & 10 & -9.10 [-17.96,-7.35] & -8.45 [-14.89,-5.86] & -7.13 [-15.62,-4.84] \\
\hdashline
Weighted & PMC (Multinomial) & 100 & 10 & -7.58 [-18.34,-5.27] & -6.46 [-16.48,-4.79] & -5.04 [-12.74,-2.93] \\
Weighted & PMC (Systematic) & 100 & 10 & -7.65 [-20.18,-5.51] & -6.58 [-19.46,-4.49] & -5.45 [-14.33,-3.25] \\
Weighted & PMC (Multinomial + Lookback) & 100 & 10 & -9.94 [-27.29,-8.11] & -9.70 [-19.04,-8.09] & -9.17 [-18.39,-6.75] \\
Weighted & PMC (Systematic + Lookback) & 100 & 10 & -9.74 [-20.70,-7.33] & -9.58 [-16.52,-7.70] & -9.16 [-16.64,-6.91] \\
Weighted & PQMC (ISP + Lookback) & 100 & 10 & \textbf{-11.78} [-20.46,-10.03] & \textbf{-11.48} [-21.30,-9.06] & \textbf{-10.93} [-21.57,-9.27] \\
\hline
  \end{tabular}
  }
  \caption{LogMSEs in the estimation of $Z$ for the two dimensional mixture of five normals using different values of $K$, $J$, and $\sigma$ with the initial proposal centers being the $K$ Sobol' points over $[0,1]^2$. The number of evaluations of the target distribution is fixed to $TKJ = 10{,}000$. The MSEs are averaged over 100 independent runs and shown in log under format ``mean [min,max]". The best results for each value of $\sigma$ are highlighted in red bold-face.}
  \label{tab:pmc_2d_full_z}
\end{table}

\begin{table}[H]
  \centering
  \resizebox{\columnwidth}{!}{%
  \begin{tabular}{|cccc|ccc|}
  \hline
  Estimator & Algorithm & K & J & $\sigma = 0.1$ & $\sigma = 0.2$ & $\sigma = 0.5$ \\
\hline
Standard & PMC (Multinomial) & 25 & 40 & -4.58 [-10.84,-2.33] & -8.43 [-13.42,-6.30] & -8.21 [-13.14,-6.68] \\
Standard & PMC (Systematic) & 25 & 40 & -4.81 [-12.55,-3.18] & -8.70 [-12.43,-7.24] & -8.13 [-13.53,-6.31] \\
Standard & PMC (Multinomial + Lookback) & 25 & 40 & -5.92 [-14.74,-3.77] & -7.98 [-15.22,-5.48] & -7.80 [-13.79,-4.05] \\
Standard & PMC (Systematic + Lookback) & 25 & 40 & -6.38 [-11.18,-4.54] & -8.75 [-13.11,-5.85] & -8.81 [-14.31,-6.19] \\
Standard & PQMC (ISP + Lookback) & 25 & 40 & -6.20 [-15.47,-4.68] & -9.98 [-14.11,-7.99] & -10.18 [-16.84,-8.55] \\
\hdashline
Weighted & PMC (Multinomial) & 25 & 40 & -4.68 [-12.03,-3.47] & -8.28 [-14.38,-6.26] & -8.22 [-13.26,-6.81] \\
Weighted & PMC (Systematic) & 25 & 40 & -4.83 [-12.14,-3.15] & -8.70 [-14.53,-6.81] & -8.13 [-11.94,-6.28] \\
Weighted & PMC (Multinomial + Lookback) & 25 & 40 & -5.94 [-15.59,-3.70] & -7.45 [-16.99,-4.92] & -7.73 [-15.16,-4.96] \\
Weighted & PMC (Systematic + Lookback) & 25 & 40 & \textbf{-6.66} [-15.00,-4.80] & -8.66 [-15.48,-5.35] & -8.33 [-17.69,-5.46] \\
Weighted & PQMC (ISP + Lookback) & 25 & 40 & -6.48 [-18.93,-4.85] & \textcolor{red}{\textbf{-14.18}} [-19.08,-12.27] & \textcolor{red}{\textbf{-14.23}} [-20.85,-12.44] \\
\hline
Standard & PMC (Multinomial) & 50 & 20 & -4.72 [-10.75,-2.80] & -8.56 [-11.71,-6.80] & -8.10 [-12.30,-5.75] \\
Standard & PMC (Systematic) & 50 & 20 & -4.82 [-14.01,-2.47] & -8.79 [-14.46,-6.78] & -8.12 [-12.81,-6.15] \\
Standard & PMC (Multinomial + Lookback) & 50 & 20 & -6.08 [-12.96,-4.57] & -9.49 [-13.96,-7.92] & -9.30 [-14.81,-7.25] \\
Standard & PMC (Systematic + Lookback) & 50 & 20 & -6.48 [-13.39,-4.27] & -9.40 [-15.52,-7.40] & -9.54 [-17.40,-7.97] \\
Standard & PQMC (ISP + Lookback) & 50 & 20 & -6.15 [-13.64,-4.71] & -9.95 [-14.40,-8.18] & -9.93 [-14.26,-8.53] \\
\hdashline
Weighted & PMC (Multinomial) & 50 & 20 & -4.88 [-13.40,-3.66] & -8.50 [-11.77,-6.71] & -8.10 [-19.25,-5.85] \\
Weighted & PMC (Systematic) & 50 & 20 & -5.01 [-13.68,-3.71] & -8.75 [-14.83,-6.67] & -8.08 [-13.41,-6.06] \\
Weighted & PMC (Multinomial + Lookback) & 50 & 20 & -6.45 [-15.09,-4.84] & -11.39 [-16.42,-9.68] & -11.37 [-16.52,-9.79] \\
Weighted & PMC (Systematic + Lookback) & 50 & 20 & \textbf{-6.99} [-16.22,-4.82] & -11.62 [-16.86,-9.54] & -11.68 [-17.21,-10.26] \\
Weighted & PQMC (ISP + Lookback) & 50 & 20 & -6.39 [-17.46,-4.86] & \textbf{-13.67} [-18.79,-12.08] & \textbf{-13.79} [-18.82,-12.33] \\
\hline
Standard & PMC (Multinomial) & 100 & 10 & -4.83 [-10.93,-2.41] & -8.83 [-14.63,-7.20] & -8.13 [-12.86,-6.51] \\
Standard & PMC (Systematic) & 100 & 10 & -4.95 [-11.41,-2.21] & -8.89 [-13.40,-7.01] & -7.91 [-12.79,-6.23] \\
Standard & PMC (Multinomial + Lookback) & 100 & 10 & -5.84 [-13.22,-2.60] & -9.57 [-17.21,-7.84] & -9.03 [-14.37,-5.43] \\
Standard & PMC (Systematic + Lookback) & 100 & 10 & -6.65 [-13.36,-4.06] & -9.60 [-14.63,-8.04] & -9.49 [-14.46,-7.71] \\
Standard & PQMC (ISP + Lookback) & 100 & 10 & -6.25 [-12.99,-2.47] & -9.68 [-13.20,-8.24] & -9.63 [-17.26,-8.46] \\
\hdashline
Weighted & PMC (Multinomial) & 100 & 10 & -5.06 [-12.29,-3.86] & -8.83 [-13.82,-6.94] & -8.11 [-12.49,-6.64] \\
Weighted & PMC (Systematic) & 100 & 10 & -5.11 [-13.59,-3.33] & -8.88 [-13.20,-7.19] & -7.87 [-15.64,-6.16] \\
Weighted & PMC (Multinomial + Lookback) & 100 & 10 & -6.34 [-18.00,-4.82] & -11.94 [-17.33,-10.32] & -9.39 [-19.47,-4.87] \\
Weighted & PMC (Systematic + Lookback) & 100 & 10 & -7.26 [-16.49,-4.81] & -12.04 [-18.51,-10.04] & -12.04 [-15.79,-10.61] \\
Weighted & PQMC (ISP + Lookback) & 100 & 10 & \textcolor{red}{\textbf{-7.28}} [-16.89,-4.86] & \textbf{-13.26} [-17.76,-11.39] & \textbf{-13.15} [-17.83,-11.40] \\
\hline
  \end{tabular}
  }
  \caption{LogMSEs in the estimation of $\mathbb{E}_{\pi}[X]$ for the two dimensional mixture of five normals using different values of $K$, $J$, and $\sigma$ with the initial proposal centers being the $K$ Sobol' points over $[0.4,0.6]^2$. The number of evaluations of the target distribution is fixed to $TKJ = 10{,}000$. The MSEs are averaged over 100 independent runs and shown in log under format ``mean [min,max]". The best results for each value of $\sigma$ are highlighted in red bold-face.}
  \label{tab:pmc_2d_sub_m}
\end{table}

\begin{table}[H]
  \centering
  \resizebox{\columnwidth}{!}{%
  \begin{tabular}{|cccc|ccc|}
  \hline
Estimator & Algorithm & K & J & $\sigma = 0.1$ & $\sigma = 0.2$ & $\sigma = 0.5$ \\
\hline
Standard & PMC (Multinomial) & 25 & 40 & 0.26 [-9.12,4.82] & -6.08 [-19.01,-4.41] & -5.75 [-13.51,-4.10] \\
Standard & PMC (Systematic) & 25 & 40 & -2.91 [-14.01,-0.65] & -6.32 [-21.09,-3.97] & -5.75 [-16.57,-3.80] \\
Standard & PMC (Multinomial + Lookback) & 25 & 40 & -3.81 [-20.37,-1.53] & -5.84 [-15.83,-3.52] & -4.62 [-18.38,-0.35] \\
Standard & PMC (Systematic + Lookback) & 25 & 40 & -4.34 [-12.79,-2.63] & -6.31 [-15.58,-3.65] & -6.22 [-16.76,-4.16] \\
Standard & PQMC (ISP + Lookback) & 25 & 40 & -4.16 [-14.93,-2.63] & -7.83 [-24.61,-5.51] & -7.60 [-13.43,-5.73] \\
\hdashline
Weighted & PMC (Multinomial) & 25 & 40 & -2.81 [-10.72,-0.98] & -5.94 [-17.44,-3.70] & -5.64 [-15.69,-3.77] \\
Weighted & PMC (Systematic) & 25 & 40 & -3.02 [-11.64,-1.05] & -6.26 [-15.95,-3.62] & -5.69 [-15.55,-3.88] \\
Weighted & PMC (Multinomial + Lookback) & 25 & 40 & -4.03 [-16.48,-1.75] & -5.35 [-19.25,-3.22] & -5.69 [-19.74,-3.10] \\
Weighted & PMC (Systematic + Lookback) & 25 & 40 & \textbf{-4.88} [-15.12,-3.08] & -6.05 [-20.75,-3.25] & -5.89 [-15.81,-3.22] \\
Weighted & PQMC (ISP + Lookback) & 25 & 40 & -4.82 [-18.64,-3.16] & \textcolor{red}{\textbf{-11.94}} [-23.84,-9.73] & \textcolor{red}{\textbf{-11.61}} [-21.25,-9.81] \\
\hline
Standard & PMC (Multinomial) & 50 & 20 & -2.13 [-15.06,1.30] & -6.27 [-17.21,-3.83] & -5.53 [-13.97,-3.61] \\
Standard & PMC (Systematic) & 50 & 20 & -2.61 [-11.38,0.89] & -6.29 [-12.02,-4.32] & -5.53 [-13.18,-3.54] \\
Standard & PMC (Multinomial + Lookback) & 50 & 20 & -3.80 [-11.55,-0.48] & -7.10 [-18.48,-5.08] & -6.93 [-19.37,-4.89] \\
Standard & PMC (Systematic + Lookback) & 50 & 20 & -4.51 [-11.91,-2.50] & -7.03 [-18.43,-5.08] & -7.13 [-20.08,-5.22] \\
Standard & PQMC (ISP + Lookback) & 50 & 20 & -3.94 [-15.83,-1.22] & -7.37 [-16.70,-4.54] & -7.15 [-19.28,-5.07] \\
\hdashline
Weighted & PMC (Multinomial) & 50 & 20 & -3.07 [-10.86,-1.01] & -6.23 [-13.08,-3.90] & -5.31 [-14.59,-3.38] \\
Weighted & PMC (Systematic) & 50 & 20 & -3.29 [-11.86,-1.87] & -6.36 [-14.31,-4.40] & -5.44 [-12.60,-3.77] \\
Weighted & PMC (Multinomial + Lookback) & 50 & 20 & -4.78 [-15.50,-3.04] & -9.00 [-23.68,-7.13] & -8.78 [-18.28,-6.53] \\
Weighted & PMC (Systematic + Lookback) & 50 & 20 & \textbf{-5.31} [-21.31,-3.11] & -9.18 [-18.06,-7.34] & -9.25 [-18.59,-7.58] \\
Weighted & PQMC (ISP + Lookback) & 50 & 20 & -4.72 [-16.52,-3.17] & \textbf{-11.10} [-23.26,-8.75] & \textbf{-11.23} [-19.51,-9.15] \\
\hline
Standard & PMC (Multinomial) & 100 & 10 & 0.26 [-9.84,4.77] & -6.50 [-21.46,-4.50] & -5.56 [-14.24,-3.71] \\
Standard & PMC (Systematic) & 100 & 10 & 5.41 [-13.55,10.01] & -6.37 [-14.26,-4.25] & -5.40 [-13.99,-3.08] \\
Standard & PMC (Multinomial + Lookback) & 100 & 10 & -2.45 [-14.07,1.96] & -7.10 [-19.66,-5.19] & -6.91 [-17.22,-3.55] \\
Standard & PMC (Systematic + Lookback) & 100 & 10 & -4.69 [-19.70,-2.60] & -7.02 [-17.75,-5.04] & -6.90 [-16.81,-4.42] \\
Standard & PQMC (ISP + Lookback) & 100 & 10 & -1.96 [-14.73,2.57] & -7.49 [-15.95,-5.70] & -7.48 [-15.27,-5.59] \\
\hdashline
Weighted & PMC (Multinomial) & 100 & 10 & -3.32 [-17.99,-1.83] & -6.41 [-13.43,-4.31] & -5.39 [-14.47,-3.52] \\
Weighted & PMC (Systematic) & 100 & 10 & -3.21 [-17.34,-0.16] & -6.33 [-15.53,-4.29] & -5.23 [-18.95,-2.81] \\
Weighted & PMC (Multinomial + Lookback) & 100 & 10 & -4.70 [-16.33,-3.05] & -9.32 [-17.47,-6.98] & -7.59 [-17.85,-3.17] \\
Weighted & PMC (Systematic + Lookback) & 100 & 10 & -5.58 [-16.53,-3.10] & -9.45 [-18.23,-7.84] & -9.35 [-17.46,-7.47] \\
Weighted & PQMC (ISP + Lookback) & 100 & 10 & \textcolor{red}{\textbf{-5.60}} [-20.48,-3.14] & \textbf{-10.61} [-20.74,-8.59] & \textbf{-10.65} [-22.69,-8.57] \\
\hline
  \end{tabular}
  }
  \caption{LogMSEs in the estimation of $Z$ for the two dimensional mixture of five normals using different values of $K$, $J$, and $\sigma$ with the initial proposal centers being the $K$ Sobol' points over $[0.4,0.6]^2$. The number of evaluations of the target distribution is fixed to $TKJ = 10{,}000$. The MSEs are averaged over 100 independent runs and shown in log under format ``mean [min,max]". The best results for each value of $\sigma$ are highlighted in red bold-face.}
  \label{tab:pmc_2d_sub_z}
\end{table}

\begin{table}[H]
  \centering
  \resizebox{\columnwidth}{!}{%
  \begin{tabular}{|cccc|ccc|}
  \hline
Estimator & Algorithm & K & J & $\sigma = 0.1$ & $\sigma = 0.2$ & $\sigma = 0.5$ \\
\hline
Standard & PMC (Multinomial) & 50 & 40 & 0.13 [-5.90,3.60] & -7.17 [-14.59,-5.00] & -2.99 [-10.89,-0.84] \\
Standard & PMC (Systematic) & 50 & 40 & -0.08 [-8.00,3.61] & -7.27 [-14.61,-4.82] & -3.36 [-11.42,-0.87] \\
Standard & PMC (Multinomial + Lookback) & 50 & 40 & -4.50 [-13.71,-1.93] & -6.96 [-16.44,-3.61] & -5.21 [-15.41,-2.14] \\
Standard & PMC (Systematic + Lookback) & 50 & 40 & -4.41 [-14.44,-1.27] & -7.17 [-18.47,-5.23] & -5.22 [-14.02,-2.39] \\
Standard & PQMC (ISP + Lookback) & 50 & 40 & -4.57 [-10.52,-2.18] & -7.47 [-12.86,-5.03] & -5.18 [-13.47,-1.90] \\
\hdashline
Weighted & PMC (Multinomial) & 50 & 40 & -1.00 [-6.63,-0.55] & -8.13 [-21.41,-6.16] & -3.03 [-10.89,-1.45] \\
Weighted & PMC (Systematic) & 50 & 40 & -1.07 [-2.01,-0.74] & -8.12 [-16.57,-6.15] & -3.01 [-10.70,-1.45] \\
Weighted & PMC (Multinomial + Lookback) & 50 & 40 & -7.54 [-15.48,-5.41] & -8.26 [-15.55,-6.18] & -7.79 [-16.44,-5.85] \\
Weighted & PMC (Systematic + Lookback) & 50 & 40 & -8.01 [-17.04,-6.20] & -8.19 [-17.52,-6.25] & -7.91 [-17.79,-6.00] \\
Weighted & PQMC (ISP + Lookback) & 50 & 40 & \textbf{-8.77} [-23.23,-7.05] & \textbf{-9.24} [-20.40,-7.14] & \textbf{-9.09} [-15.29,-7.23] \\
\hline
Standard & PMC (Multinomial) & 100 & 20 & -1.07 [-6.83,2.45] & -7.57 [-16.99,-5.29] & -2.84 [-12.40,0.19] \\
Standard & PMC (Systematic) & 100 & 20 & -0.89 [-10.74,2.90] & -7.73 [-16.26,-5.66] & -2.95 [-14.81,-0.84] \\
Standard & PMC (Multinomial + Lookback) & 100 & 20 & 0.32 [-15.17,4.89] & -7.56 [-18.75,-5.39] & -4.38 [-15.43,-0.99] \\
Standard & PMC (Systematic + Lookback) & 100 & 20 & -4.20 [-18.48,-0.44] & -7.63 [-16.37,-5.55] & -4.93 [-13.44,-2.17] \\
Standard & PQMC (ISP + Lookback) & 100 & 20 & -5.07 [-12.55,-4.04] & -7.94 [-19.68,-5.08] & -5.44 [-14.37,-2.53] \\
\hdashline
Weighted & PMC (Multinomial) & 100 & 20 & -1.31 [-4.95,-0.89] & -8.30 [-25.44,-6.32] & -2.90 [-13.24,-1.64] \\
Weighted & PMC (Systematic) & 100 & 20 & -1.46 [-2.75,-1.05] & -8.67 [-16.82,-6.96] & -3.00 [-10.95,-1.44] \\
Weighted & PMC (Multinomial + Lookback) & 100 & 20 & -8.40 [-20.16,-6.09] & -8.33 [-22.86,-6.57] & -7.85 [-16.30,-5.41] \\
Weighted & PMC (Systematic + Lookback) & 100 & 20 & -8.05 [-17.03,-6.08] & -8.57 [-14.54,-7.03] & -8.64 [-16.32,-6.68] \\
Weighted & PQMC (ISP + Lookback) & 100 & 20 & \textcolor{red}{\textbf{-9.09}} [-19.73,-7.01] & \textcolor{red}{\textbf{-9.28}} [-19.36,-6.99] & \textcolor{red}{\textbf{-9.18}} [-19.98,-6.76] \\
\hline
  \end{tabular}
  }
  \caption{LogMSEs in the estimation of $Z$ for the ten dimensional mixture of three normals using different values of $K$, $J$, and $\sigma$ with the initial proposal centers being the $K$ Sobol' points over $[0,1]^{10}$. The number of evaluations of the target distribution is fixed to $TKJ = 20{,}000$. The MSEs are averaged over 100 independent runs and shown in log under format ``mean [min,max]". The best results for each value of $\sigma$ are highlighted in red bold-face.}
  \label{tab:pmc_10d_full_z}
\end{table}

\begin{figure}[H]
  \centering
  
  \begin{subfigure}{\textwidth}
    \centering
    \includegraphics[width=\textwidth]{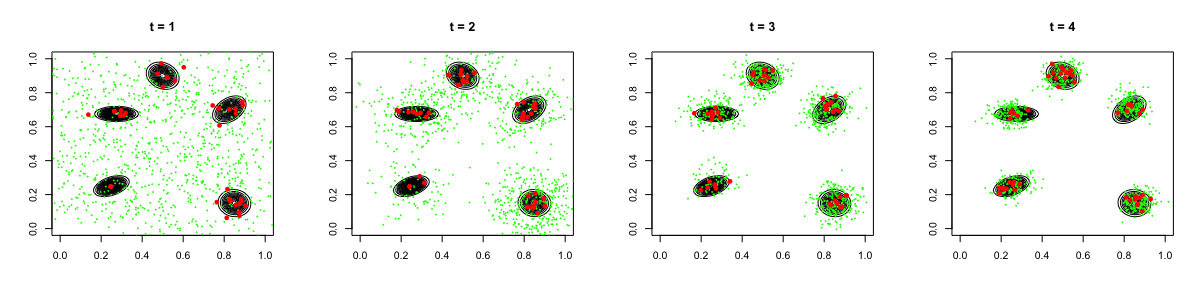}
    \caption{PMC (Multinomial + Lookback)}
  \end{subfigure}%
  
  \begin{subfigure}{\textwidth}
    \centering
    \includegraphics[width=\textwidth]{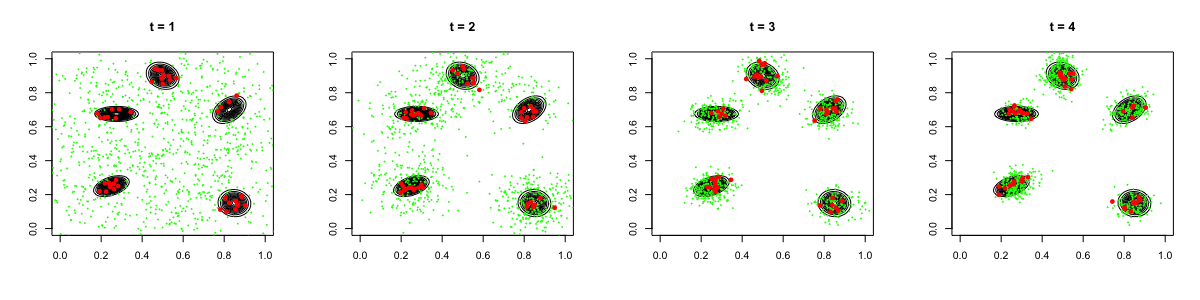}
    \caption{PMC (Systematic + Lookback)}
  \end{subfigure}%
  
  \begin{subfigure}{\textwidth}
    \centering
    \includegraphics[width=\textwidth]{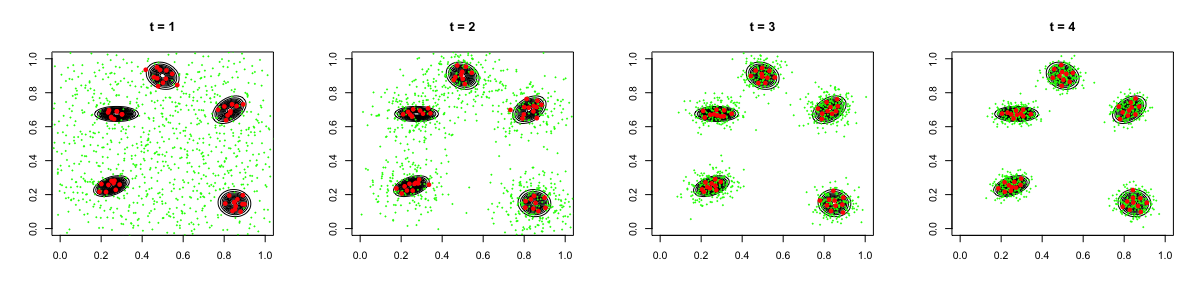}
    \caption{PQMC (ISP + Lookback)}
  \end{subfigure}%
  
  \caption{Evolution of the weighted samples (green diamonds) and the resamples (red dots) for the first 4 iterations of both PMC and PQMC using $K = 50$, $J = 20$, and $\sigma = 0.1$ on the mixture of five normals example (Subsection~\ref{subsec:simulation_pmc_2d}). The initial centers are the 50 Sobol' points over $[0,1]^2$. Lines represent the density contours.}
  \label{fig:pmc_2d}
\end{figure}

\begin{figure}[H]
  \centering
  
  \begin{subfigure}{0.9\textwidth}
    \centering
    \includegraphics[width=\textwidth]{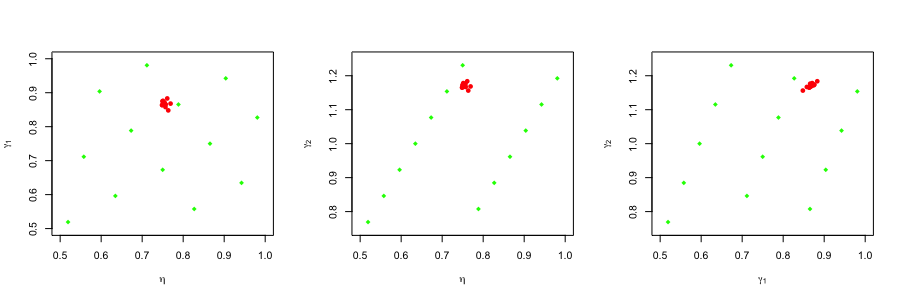}
    \caption{PMC (Multinomial + Lookback)}
  \end{subfigure}%
  
  \begin{subfigure}{0.9\textwidth}
    \centering
    \includegraphics[width=\textwidth]{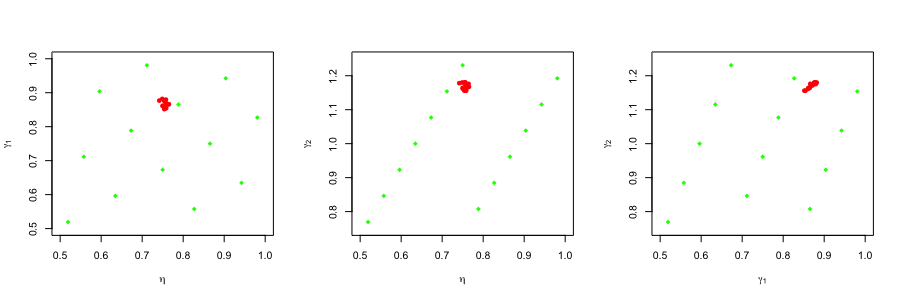}
    \caption{PMC (Systematic + Lookback)}
  \end{subfigure}%
  
  \begin{subfigure}{\textwidth}
    \centering
    \includegraphics[width=0.9\textwidth]{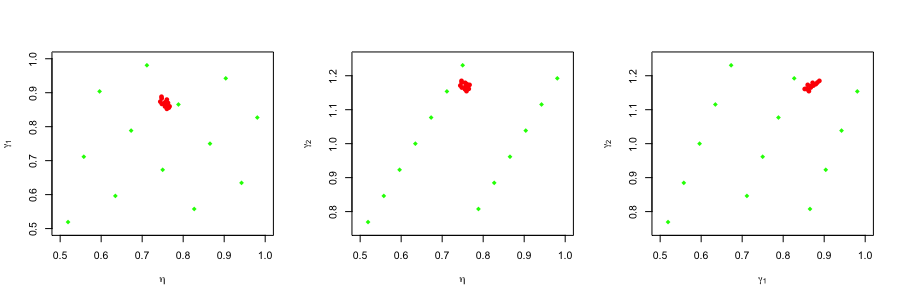}
    \caption{PQMC (ISP + Lookback)}
  \end{subfigure}%
  
  \caption{Initial (green diamonds) and final (red dots) 13 proposal centers for the PMC and PQMC algorithms on the friction drilling example (Section~\ref{sec:simulation_pmc_drilling}).}
  \label{fig:drilling_pp}
\end{figure}

\end{document}